\documentclass[12pt]{emulateapj}
\slugcomment{Accepted, to appear in ApJ, v701, 2009 August.}
\pdfoutput=1
\usepackage{natbib}
\newcommand{\gtorder}{\mathrel{\raise.3ex\hbox{$>$}\mkern-14mu
             \lower0.6ex\hbox{$\sim$}}}
\newcommand{\ltorder}{\mathrel{\raise.3ex\hbox{$<$}\mkern-14mu
             \lower0.6ex\hbox{$\sim$}}}

\newcommand\eg{{\it e.g.,~}}
\newcommand\ie{{\it i.e.,~}}

\begin{document}
\begin{abstract}

The association of an electromagnetic signal with the merger of a
pair of supermassive black holes would have many important
implications.  For example, it would provide new information about
gas and magnetic field interactions in dynamical spacetimes as well
as a combination of redshift and luminosity distance that
would enable precise cosmological tests.  A proposal first made by
\citet{2007APS..APR.S1010B} is that because radiation of gravitational waves
during the final inspiral and merger of the holes is abrupt and
decreases the mass of the central object by a few percent, there
will be waves in the disk that can steepen into shocks and thus
increase the disk luminosity in a characteristic way.  We evaluate
this process analytically and numerically.  We find that shocks
only occur when the fractional mass loss exceeds the half-thickness ($h/r$)
of the disk, hence significant energy release only occurs for 
geometrically thin disks which are thus at low Eddington ratios.  
This strongly limits the effective energy release, and in fact our
simulations show that the natural variations in
disk luminosity are likely to obscure this effect entirely. 
However, we demonstrate that the {\it reduction} of luminosity caused by the
retreat of the inner edge of the disk following mass loss is
potentially detectable.  This decrease occurs even if the disk is
geometrically thick, and lasts for a duration on the order of the viscous time
of the modified disk.  Observationally, the best prospect for
detection would be a sensitive future X-ray instrument with a field
of view of on the order of a square degree, or possibly a
wide-field radio array such as the Square Kilometer Array, if the
disk changes produce or interrupt radio emission from a jet.

\end{abstract}


\title{Reaction of Accretion Disks to Abrupt Mass Loss During Binary
Black Hole Merger}
\author{Sean M. O'Neill\altaffilmark{1}, M. Coleman Miller\altaffilmark{1}, Tamara Bogdanovi\'c\altaffilmark{1},
\\
Christopher S. Reynolds\altaffilmark{1}, and Jeremy D. Schnittman\altaffilmark{2}}
\altaffiltext{1}{University of Maryland, Department of Astronomy and
Maryland Astronomy Center for Theory and Computation,\\
College Park, Maryland 20742-2421}
\altaffiltext{2}{Johns Hopkins University}
\maketitle

\section{Introduction}

There has recently been significant interest in whether the
coalescence of two supermassive black holes can produce an
identifiable electromagnetic (EM) signature.  If a redshift could be
extracted from such a signature (via localization of the host
galaxy), then it would provide a powerful cosmological probe
when combined with the expected estimate of the luminosity
distance, which will have a precision of $\sim 10$\% mainly
limited by weak gravitational lensing \citep{2003CQGra..20S..65H}.  It would
also allow precise tests of whether gravitons travel at the
speed of light, as required by general relativity.

While the merger itself produces no electromagnetic emission,
if there are significant electromagnetic fields or mass
nearby in an accretion disk then there are various
possibilities.  \cite{2005ApJ...622L..93M} point out that for
some disk accretion rates and binary mass ratios, the binary
reaches a point in its coalescence such that further inspiral by emission of gravitational waves
occurs more rapidly than the disk diffuses inwards.  This leads
to a hole in the disk which is filled gradually after merger,
leading to a source that brightens over weeks to years depending
on various parameters. Several recent authors
\citep{2008ApJ...682..758S,2008ApJ...676L...5L,2008ApJ...684..835S} 
discuss consequences of the recoils from asymmetric emission of
gravitational waves during the coalescence, from prompt shocks to
delayed emission lasting millions of years.  Emission might occur
in the late inspiral 
\citep{2002ApJ...567L...9A,2008PhRvL.101d1101K,2008arXiv0811.1920H}
from effects such as enhanced accretion, periodic Newtonian 
perturbations, or shearing of the disk due to gravitational waves.
Earlier precursors are also possible
\citep{2006MNRAS.372..869D,2008ApJ...684..870K,2009arXiv0904.1383H},
and in some cases the error volume from the gravitational wave
signal may be small enough that the host galaxy can be identified
by its morphology or mass, or by the presence of an Active Galactic Nucleus (AGN) (e.g.,
\citealt{2006ApJ...637...27K}).

A suggestion first made by \cite{2007APS..APR.S1010B} is that the effectively
instantaneous mass-energy loss to gravitational radiation would
leave the disk in a non-equilibrium state, and that adjustment of
the disk could produce observable radiation. Their specific
suggestion was that because fluid orbits would be elliptical after
mass loss, circularization would release extra energy that could
be detected.   In this paper, we examine this scenario and find
that in most circumstances this energy release is too small, and
the timescale too large, for such emission to be realistically
detectable.  We show that in principle, for systems accreting at a
high enough rate that the inner edge can stay within a factor of
$\sim 2$ of the binary semimajor axis all the way down to binary
dynamical instability (\ie when the black holes begin to plunge into one another), relativistic effects enhance the energy
release significantly.
Rapid release of this energy, however, requires the development of shocks.
As we demonstrate, shock formation only occurs if the disk is sufficiently geometrically thin.  
Such disks are characterized
by a long viscous timescale and thus are not sufficiently
relativistic at the point of merger for there to be substantial
energy release due to circularization.

Here we show that a stronger signal may be obtained from the
prompt {\it drop} in emission that occurs because the inner edge
of the disk retreats after mass loss and then takes some time
to fill in.  The resulting flux deficit compared to the emission
prior to mass loss is in the detectable range for current 
instruments such as {\it Chandra}, and observable for
planned observatories such as the {\it International X-ray
Observatory}, if the angular localization of the gravitational
wave event prior to merger is sufficiently precise.
In \S~2 we
discuss the basic energetics and timescales for both Newtonian and 
relativistic disks.  In \S~3 we give the results of our hydrodynamic
and magnetohydrodynamic simulations.  In \S~4 we discuss the
potential observability of this effect, particularly when measured
against the backdrop of galaxy mergers and AGN variability, and in
\S~5 we give our conclusions.

\section{Basic energetics for Newtonian and relativistic disks}

\subsection{Fractional mass loss}

Our starting point is the typical fraction of total binary mass
lost to gravitational radiation during a coalescence.  
\citet{2008arXiv0807.2985T} analyze the results of numerous
simulations and propose an empirical relation that accurately estimates $\Delta m/M_{\rm tot}$, where $\Delta m$ is
the mass-equivalent lost to radiation and $M_{\rm tot}$ was the
mass of the binary when the black holes were well separated.  They
find
\begin{equation}
\Delta m/M_{\rm tot}\approx 0.0485(4\eta)+0.013(a_z+b_z)(4\eta)^2\; .
\end{equation}
Here, for a black hole binary of masses $m_a$ and $m_b$, $\eta$ is the
symmetric mass ratio $\eta\equiv m_am_b/(m_a+m_b)^2$.  This has
a maximum of $\eta_{\rm max}=1/4$ for an equal-mass binary.
We also define $a_z$ and $b_z$ to be the dimensionless angular momenta
of the two holes in the direction of the orbital angular
momentum, $a_z=cJ_{a,z}/(Gm_a^2)$ and $b_z=cJ_{b,z}/(Gm_b^2)$.

For gas-rich mergers such as those likely to produce the
strongest electromagnetic signatures, \citet{2007ApJ...661L.147B}
argue that torques from the gas would tend to align the spins
of the individual holes with the orbital axis and hence with
each other.  Accretion is thus usually prograde, hence black holes 
in gas-rich environments accreting for extended times are expected
to spin rapidly; observational evidence for high spins has
been presented via analysis of iron K$\alpha$ lines
(e.g., \citealt{1996MNRAS.282.1038I,2002MNRAS.335L...1F,2003PhR...377..389R,2006ApJ...652.1028B}).
For these objects $a_z\approx b_z\approx 1$ and the mass loss is
maximized.  A reasonable approximation to the
fractional mass loss is therefore
$\Delta m/M_{\rm tot}\approx 0.05(4\eta)+0.025(4\eta)^2$.
This is roughly 7.5\% for equal-mass systems, 5\% for
a 1:3 mass ratio, and 2\% for a 1:10 mass ratio.

A useful simplification is that most of the mass is lost at
the very end of coalescence, hence from the perspective of
the disk there is a rapid reduction of the mass of
the central object.  For example, when two equal-mass nonspinning
black holes coalesce, roughly 70\% of the total energy radiated
comes out in the final orbit plus merger and ringdown (e.g., \citealt{2005PhRvL..95l1101P} and numerous
subsequent papers).  We will therefore treat all
such losses as instantaneous.
In addition, we consider the effects of mass loss to be decoupled from
the gravitational wave kick that a remnant black hole may receive
after coalescence.  This is justified since the orbital speed in the
innermost part of the disk is large compared to the velocity of the kicked black hole.  Consequently, any EM signatures
associated with the kick occur primarily at much larger radii, and hence timescales, than the inner regions of interest here.

\subsection{Newtonian energetics and timescales}

For simplicity, we will consider first a disk with an inner edge
far enough from the merging binary that a
Newtonian treatment around a point mass is justified.  Suppose
that the fractional mass loss is $\epsilon\ll 1$, i.e., the
mass after merger is $M=M_0(1-\epsilon)$ where $M_0$ is the
initial binary mass.  We also presume that
initially the disk gas moves in essentially circular orbits,
meaning that the inward radial speed is much smaller than the
orbital speed.  Our final assumption is that the disk is 
initially axisymmetric as opposed to having significant
azimuthal variations due to spiral density waves induced by
the binary.

A fluid element at radius $r$ then has initial specific angular
momentum $\ell_0=\sqrt{GM_0 r}$.  After reduction of the central
mass, the specific angular momentum is unchanged but the fluid
element is now at the pericenter of its orbit.  If we now consider
fluid elements all at the same radius as constituting an annulus,
then the annulus will oscillate in phase.  Other
annuli will do the same thing but at different frequencies,
meaning that to lowest order the accelerations on an annulus due
to other annuli will all be radial and thus not change its
specific angular momentum. We therefore assume that the specific
angular momentum in a given fluid element is a constant and then minimize energy
under this constraint.

The initial specific energy after mass loss is $E_{\rm tot}={1\over 2}v^2-GM/r=GM_0/(2r)-GM_0(1-\epsilon)/r$.
Rewriting, $E_{\rm tot}=-GM_0(1-2\epsilon)/(2r)$.  
The initial semimajor axis is then given by $E_{\rm tot}=-GM/(2a)=-GM_0(1-\epsilon)/(2a)$.
Equating these two expressions for $E_{\rm tot}$ and keeping only the lowest-order terms, $r\approx (1-\epsilon)a$.
Thus, the eccentricity is $e\approx \epsilon$.
The specific angular momentum of an orbit with eccentricity $e$ is $l(e) = \sqrt{GMa(1-e^2)}$.
The change in binding energy as a result of orbital circularization is given by $\Delta E = [l(e)^2-l(0)^2]/(2a^2) = - GMe^2/(2a)$, so the fractional change in the binding energy becomes $\Delta E/E_{\rm tot}=e^2=\epsilon^2$.

The timescale on which this energy is released is unlikely to
be shorter than the time needed for neighboring annuli to
develop significant relative radial motion.  Orbits with
semimajor axes differing by a factor $\sim (1+e)$ will
intersect each other when their radial epicyclic oscillations
get out of phase by $\sim 1$~radian.  Since in Newtonian
gravity the radial epicyclic frequency $\kappa$ equals the 
orbital frequency, the minimum timescale at radius $r$ becomes
\begin{equation}
T_{\rm min}(r)={1\over{2\pi}}\left(er \frac{d\kappa}{dr}\right)^{-1}\approx
{0.1\over\epsilon}P_{\rm orb}(r)
\end{equation}
where $P_{\rm orb}(r)$ is the orbital period at $r$.  The actual
time of energy release could be significantly greater than
$T_{\rm min}$ if, for example, no shocks ever develop or there
is a delay in radiating the energy from the disk due to large
vertical optical depths \citep{2008ApJ...684..835S}.  This in turn implies a maximum
specific luminosity of
\begin{equation}
L_{\rm max}=\Delta E/T_{\rm min}
\approx 10\epsilon^3E_{\rm tot}/P_{\rm orb}(r)\; .
\end{equation}
This is to be compared with the natural energy release of the
disk as it flows inwards.  For example, consider a fairly large
mass loss of $\epsilon=0.05$.  Then $L_{\rm max}\approx 
10^{-3}E_{\rm tot}/P_{\rm orb}(r)$.  
For a Shakura-Sunyaev \citep{1973A&A....24..337S} viscosity
parameter $\alpha=0.1$ and disk half-thickness $h/r=0.1$,
classical disk theory predicts an inward radial speed of 
$v_r={3\over 2}\alpha (h/r)^2v_{\rm orb}$, where $v_{\rm orb}$ is the
orbital speed.  The characteristic time to
move from a radius $r$ to a radius $r/2$ is thus roughly
$P_{\rm orb}(r)/[6\pi\alpha (h/r)^2]\approx 100P_{\rm orb}(r)$, so
the specific luminosity is $L\sim 10^{-2}E_{\rm tot}/P_{\rm orb}(r)$.  The
energy release from mass loss would in this case be only a
$\sim 10$\% perturbation.
Physically, of course, it is the value of $h/r$ that is determined by the luminosity of the disk and not vice-versa, but it is still convenient to parametrize the luminosity in terms of the aspect ratio.
While this estimate of a $\sim 10 \%$ perturbation applies to a steady disk with the given parameters, we further note that disks with lower values of $\alpha$ and $h/r$ would feature even smaller perturbations.

\subsection{Relativistic energetics}

If the inward radial speed of the disk is sufficient to keep
up with the inspiraling binary, then near merger the
disk experiences relativistic dynamics.  As we discuss below,
it is unlikely that in such a case shocks would be produced
after the mass loss, because rapid inward motion requires a
thick disk with $h/r\sim 1$, hence the disk sound speed, $c_s \sim (h/r)v_\phi$, is
significantly higher than the radial oscillation speed, $v_r \sim \epsilon v_\phi$, after
merger.  Nonetheless, for completeness we show that if somehow
angular momentum transport is strongly enhanced (thereby allowing a thin disk to follow the binary all the way to merger), the resulting
energy release could be boosted by large factors.

A key feature of relativistic dynamics
is that the specific angular momentum of circular
orbits reaches a minimum, at the innermost stable circular
orbit (ISCO).  The specific angular momentum curve is thus
nearly flat near the ISCO, hence the sudden loss of mass at
merger can lead to much greater motion of the accreting gas
than in the Newtonian case.  This in turn produces larger and
potentially faster energy release, thus possibly greater
observability.

The spacetime will not be Schwarzschild, both because the
source of gravity is actually a binary and because rapid
rotation introduces frame-dragging.  Nonetheless, the ISCO
is a general feature and therefore we present a calculation
in the Schwarzschild spacetime as representative of the
expected effects.

The specific angular
momentum $\ell$ of a particle in a circular orbit with a
radius $r=xM$ in Schwarzschild coordinates around a mass $M$
(where here and henceforth we use geometrized units in
which $G=c\equiv 1$)
is given by $\ell^2=x^2M^2/(x-3)$.  The specific energy of
this particle is $E=(x-2)/\sqrt{x(x-3)}$, but more generally
for a particle of angular momentum $\ell$ in instantaneously
azimuthal motion the specific energy is 
$E=\sqrt{(1-2/x)[1+\ell^2/(x^2M^2)]}$.  

Now suppose that a fraction $\epsilon$ of the mass is taken
away instantly.  As before, we argue that the angular momentum
is conserved, so that
\begin{equation}
{x^2\over{x-3}}={x^{\prime 2}(1-\epsilon)^2\over{x^\prime -3}}
\end{equation}
where we define $x^\prime\equiv r^\prime/M$, where $r^\prime$
is the radius after circularization and $M$ is the final mass
of the merged black hole.
The specific energy immediately after mass loss can then
be rewritten as
\begin{equation}
E_{\rm init}=\sqrt{(x-2)[x-2(1-\epsilon)]\over{x(x-3)}}
\end{equation}
and the final specific energy after circularization is
\begin{equation}
E_{\rm fin}={x^\prime-2\over{\sqrt{x^\prime(x^\prime-3)}}}.
\end{equation}

These expressions include the rest-mass energy of the fluid, so to calculate the fractional change in binding energy, we use
\begin{equation}
\Delta E/E_{\rm tot}=\frac{(E_{\rm init}-E_{\rm fin})}{(1-E_{\rm init})}\; .
\end{equation}

As an example of the degree of enhancement achieved relative
to the Newtonian limit, consider a fractional mass loss
$\epsilon=0.05$.  In the Newtonian limit, $\Delta E/E_{\rm tot}=0.0025$.
For $x=6$ (the location of the ISCO), $\Delta E/E_{\rm tot}=0.0679$, or
27 times larger.  For $x=7$, $\Delta E/E_{\rm tot}=0.0295$.  The relative
enhancement is even greater for smaller fractional mass
losses.  For $\epsilon=0.01$, $\Delta E/E_{\rm tot}=0.0001$ in the Newtonian
limit.  In contrast, $\Delta E/E_{\rm tot}=0.00582$ for $x=6$ and
$\Delta E/E_{\rm tot}=0.00131$ for $x=7$.  The energy release is
tremendously enhanced by proximity to the ISCO.

The orbit-crossing timescales can also be reduced simply because the gradient
of the radial epicyclic frequency is much greater near the
ISCO; indeed, at the ISCO itself, $\kappa=0$.  We expect this
speedup, as well as the enhancement of energy release,
to apply for the real spacetime as well as for Schwarzschild, although
of course the exact numbers will be different.

Although these analytic expectations set the basic energetic stage,
numerical simulations are essential to evaluate the details.  
In particular, the preceding analytic arguments did not address how fluid quantities, such as the pressure, would react to and modify the induced epicyclic motion and the eventual energy dissipation.
In the next section we therefore present the results of our hydrodynamic
and magnetohydrodynamic (MHD) simulations.  We find that, as 
expected, in our models shocks and thus efficient
thermalization of energy only occur for disks that are geometrically
thin enough that the sound speed is less than the speed of radial
epicyclic oscillation.  This limits the accretion rate and also
means that the disk viscous time is long.  Therefore, the binary
decouples from the disk when the disk is still far from the ISCO, so
the relativistic energy enhancements do not apply.  In addition, we
find in our simulations that any luminosity enhancement is localized and minor.  
We do find, however, that the increase in disk inner radius caused by the abrupt mass loss results in a large decrease in luminosity that is potentially detectable by
future instruments.  We focus on this effect in \S~4.

\section{Hydrodynamic and MHD simulations}

Our primary goal in conducting these simulations is to examine the extent to which shocks induced from a binary merger appreciably heat the circumbinary disk or generate other potentially observable electromagnetic signatures.
It is reasonable to assume that any circumbinary disk will have evolved into a turbulent state, so we therefore begin by simulating a baseline 3D MHD disk using the methods ($\S$~\ref{sec:numerical}) and initial parameters ($\S$~\ref{sec:parameters}) described below.
In the baseline model, accretion is driven naturally by turbulence associated with the magnetorotational instability \citep[MRI,][]{1991ApJ...376..214B}, producing a disk that extends down to the inner regions of the potential near the ISCO.
As discussed in the previous section, it is important to model this inner disk since that is where both the energy enhancement and timescale are optimal for generating a detectable merger signal.
We then use the output of this baseline model to seed the initial conditions in our various mass-loss models, which are described in detail in Section \ref{sec:parameters}.

\subsection{Numerical Methods}\label{sec:numerical}
To simulate the response of a circumbinary accretion disk to an instantaneous central mass loss event, we employ the ZEUS-MP code (version 2), the details of which are described in \citet{1992ApJS...80..753S,1992ApJS...80..791S}, \citet{1992ApJS...80..819S}, and \citet{2006ApJS..165..188H}.
This code uses an Eulerian finite difference scheme to solve to second-order spatial accuracy the equations of ideal compressible fluid dynamics.
We conduct our simulations using spherical coordinates ($R, \theta, \phi$) in both full 3D MHD and ``2.5D'' hydrodynamic modes, the latter of which includes axisymmetric azimuthal velocities.
The simulations feature a polytropic equation of state ($p \propto \rho^\gamma$) with $\gamma = 5/3$.
Additionally, we include a protection routine to impose a density floor $\rho_{\rm min}$ at a value $10^{-7}$ times that of the initial midplane density.
We have also incorporated into the code an Alfv\'en speed limiter of the type described in \citet{2000ApJ...534..398M}.
This algorithm mimics the effect of the displacement current in the MHD simulations, naturally preventing the Alfv\'en speed from exceeding the speed of light.

Gravity in our simulations is handled through the inclusion of a point mass at the computational grid origin, and self-gravity of the disk is not modeled.
Since most of the gravitational wave emission in such a merger takes place in the last half-orbit, the binary is sufficiently close that it is reasonable to treat the pre-merger system as a single point mass.
This approach does, however, preclude our simulations from experiencing any of the natural effects of a true binary potential, such as the production of spiral density waves in the disk.
We assume such phenomena will be of minor importance in the overall behavior of the system.
We have modified the code to make the gravitational potential pseudo-Newtonian in form, as described by \citet{1980A&A....88...23P}:
\begin{equation}
\label{eq:pnpot}
\Phi=-\frac{GM}{R-2r_{\rm g}}, \qquad r_{\rm g} \equiv \frac{GM}{c^2}.
\end{equation}
This form of the potential features an ISCO at $R=6~{\rm M}$, the presence of which naturally augments the energy release from a binary merger (see Section 2.3).
Due to the relatively short timescale over which gravitational waves are emitted, the merger event itself is modeled as a discontinuous change of the central point mass.

All of our simulations span a radial domain $R \in (4{\rm M},100{\rm M})$.
This range is occupied by 768 zones for which $\Delta R$ increases logarithmically, with $\Delta R_{\rm min} = 0.02~{\rm M}$ at the inner radial boundary.
Both the inner and outer radial boundaries are used to enforce zero-gradient outflow.
This is accomplished by setting fluid variables in the boundaries to their values in adjacent grid zones, with the additional condition that radial velocities are restricted so that flow on to the grid from the boundaries is not allowed.
In the $\theta$ direction, resolution is concentrated in the inner seven disk scale heights on either side of the midplane.
There, $\Delta \theta \approx 2.1 \times 10^{-3}$ radians, while $\Delta \theta$ increases logarithmically outside of this region so that the total grid spans $\theta \in (0.05 \pi,0.95\pi)$ with 256 zones.
A reflecting boundary is employed in $\theta$ due to the proximity of the boundary to the coordinate pole, which would naturally reflect any approximately axisymmetric flow.
This non-uniform grid structure has the advantage of reducing the potentially unphysical influence of the outer grid boundaries by placing them physically far away from the regions of interest.
For those simulations that are fully three-dimensional, $\phi \in (0,\pi/6)$ with a uniform $\Delta \phi \approx 0.016$ radians.
This should be sufficient to allow fully three-dimensional MHD behaviors while avoiding the excessive computational cost of spanning a full $2\pi$ in azimuth.
In both the axisymmetric and 3D simulations, periodic boundaries are enforced in the $\phi$ direction.

\subsection{Model Parameters and Initialization}\label{sec:parameters}

As mentioned above, we conduct a single 3D MHD baseline simulation to generate the initial conditions for all of our models.
This simulation is run for $\sim 6160~{\rm M}$, corresponding to $\sim 100~T_{\rm ISCO}$, where $T_{\rm ISCO}$ is the orbital period at the ISCO.
Since we are interested primarily in the disk properties just before and following binary merger, we define $t=0$ as the time of the mass-loss event.
In this timescale, all data shown for reference at times $t \le 0$ reflect the output of the baseline simulation, while all data for which $t > 0$ describe responses of our various disk models to the mass-loss event.

\begin{deluxetable}{cccc}
\label{table}
\tabletypesize{\scriptsize}
\tablecaption{Simulations of Circumbinary Accretion Disks \\ Experiencing a Central Mass Loss}
\tablewidth{0.48\textwidth}
\tablehead{
\colhead{ID} &
\colhead{Type of Simulation \tablenotemark{1}} &
\colhead{Resolution \tablenotemark{2} } &
\colhead{$\epsilon$ \tablenotemark{3}} \\ }
\startdata
H10 & Hydrodynamic & $768 \times 256$      & 0.10 \\
M10 & MHD          & $768 \times 256 \times 32$ &  0.10\\
H5  & Hydrodynamic & $768 \times 256$      &  0.05\\
H1  & Hydrodynamic & $768 \times 256$      &  0.01\\
M1  & MHD          & $768 \times 256 \times 32$ &  0.01\\
\enddata
\tablenotetext{1}{All simulations are initialized from a single 3D MHD model.  Models M10 and M1 use the full data as initial conditions while models H10, H5, and H1 input only the azimuthally averaged hydrodynamic variables.}
\tablenotetext{2}{Resolution is listed as $R \times \theta$ ($\times \phi$).  The physical extent of the grid is $R \in (4{\rm M},100{\rm M})$, $\theta \in (0.05 \pi,0.95\pi)$, $\phi \in (0,\pi/6)$.}
\tablenotetext{3}{We define $\epsilon$ as the fraction of the total initial mass lost through the binary merger, and we model the central mass loss as an instantaneous event.  In the figures and discussion, the mass loss is described as taking place at a time $t = 0$.}
\end{deluxetable}

\begin{figure*}[t]
\includegraphics[type=pdf,ext=.pdf,read=.pdf,width=0.5\textwidth]{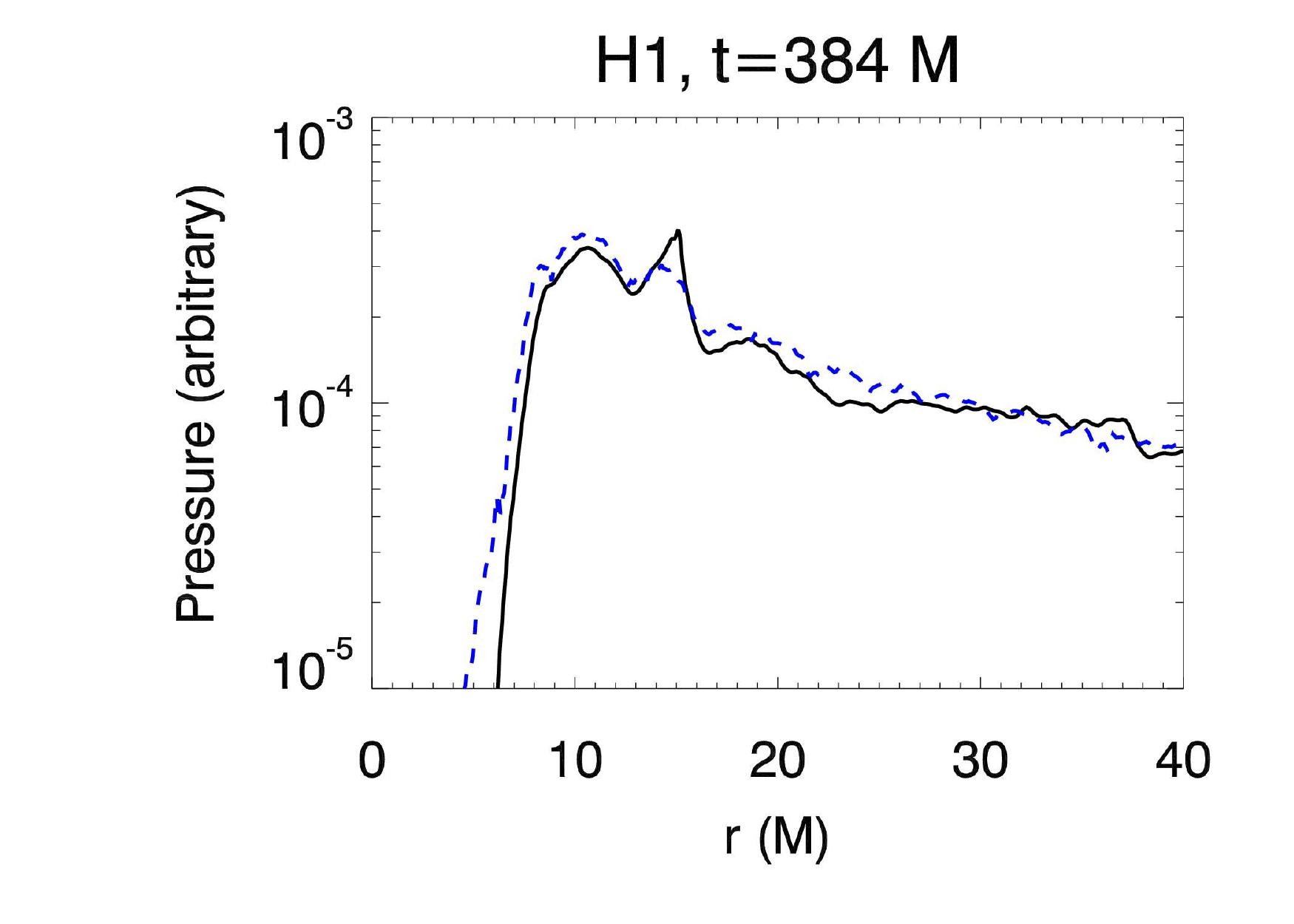}
\includegraphics[type=pdf,ext=.pdf,read=.pdf,width=0.5\textwidth]{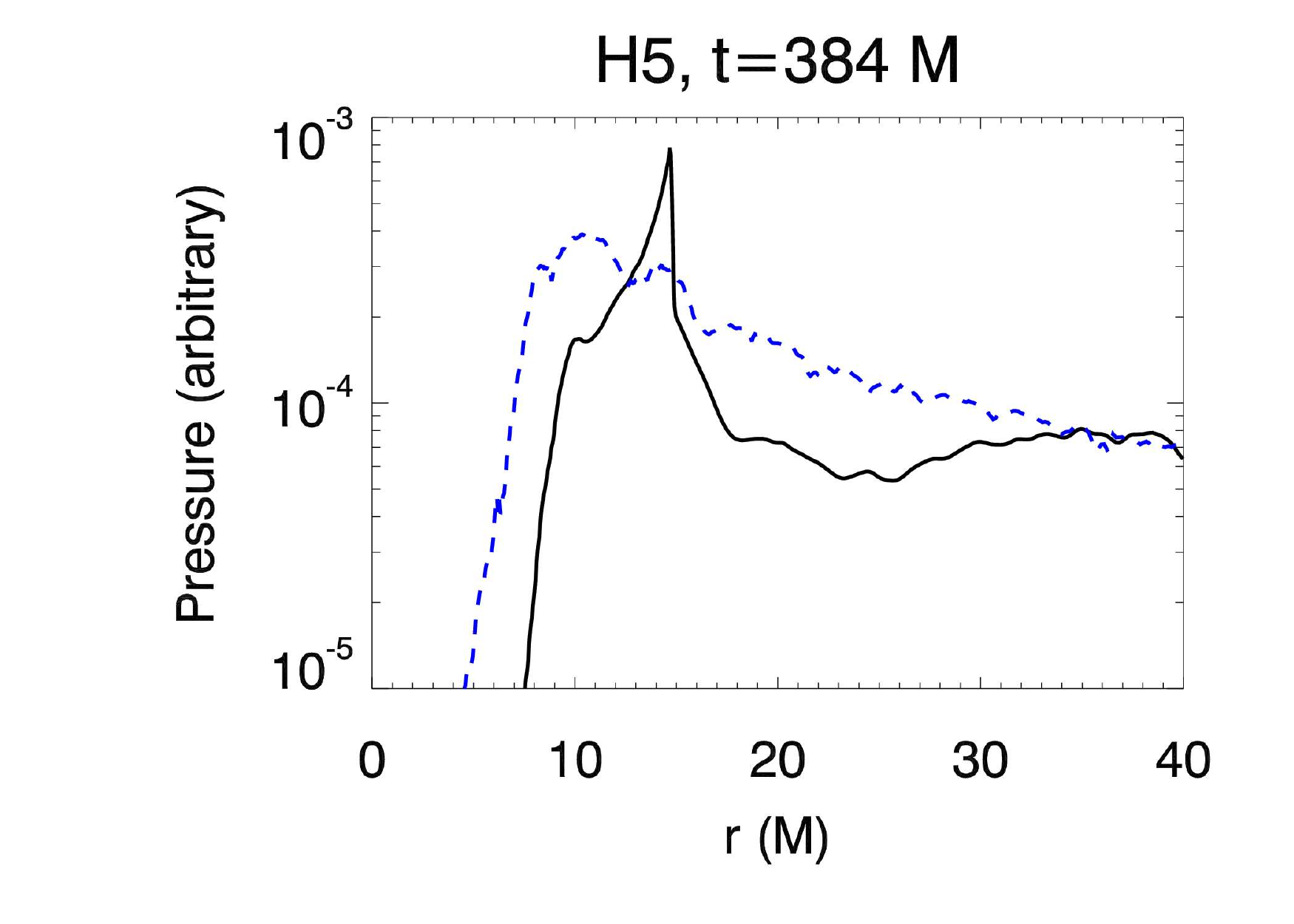}
\\
\includegraphics[type=pdf,ext=.pdf,read=.pdf,width=0.5\textwidth]{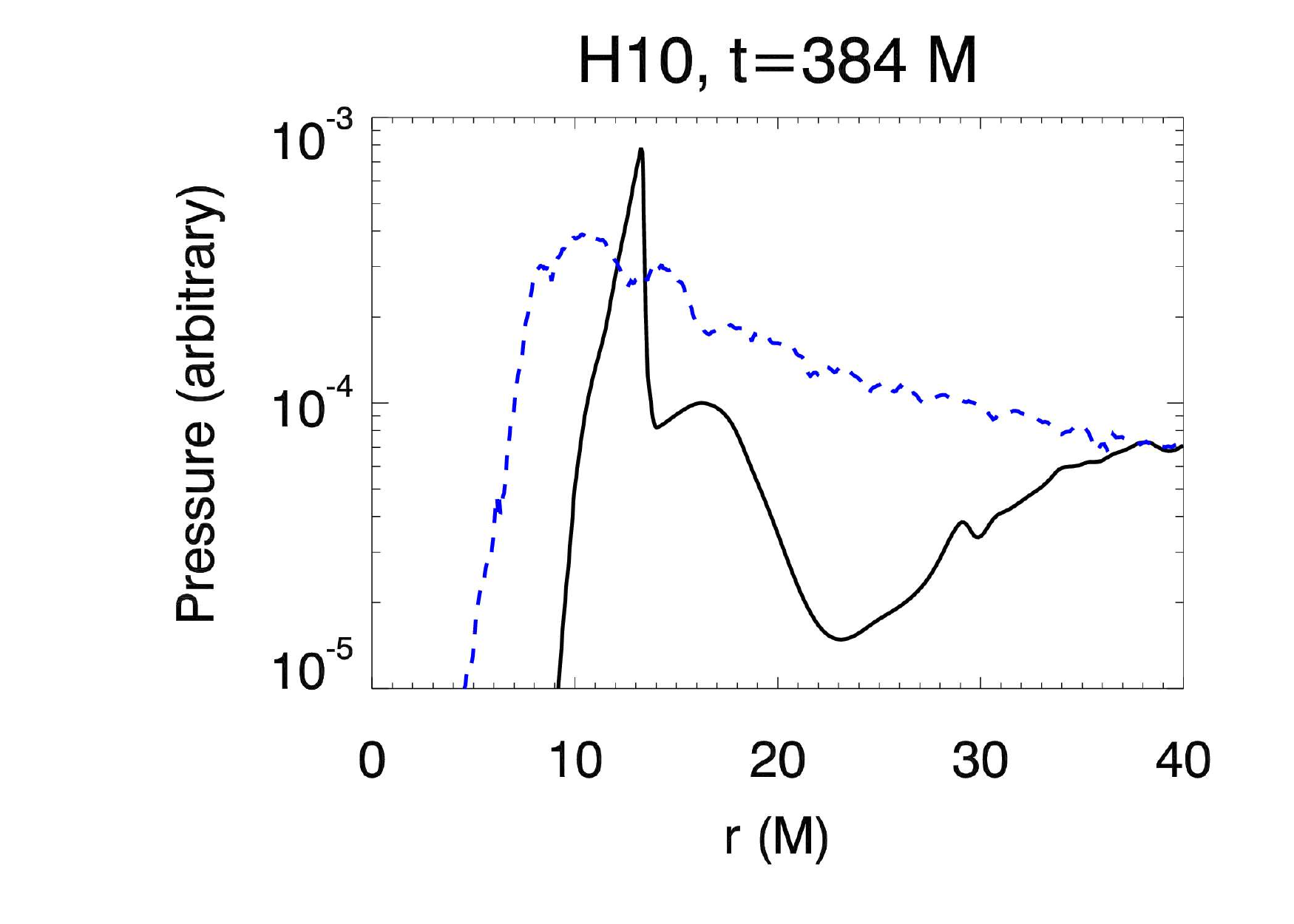}
\includegraphics[type=pdf,ext=.pdf,read=.pdf,width=0.5\textwidth]{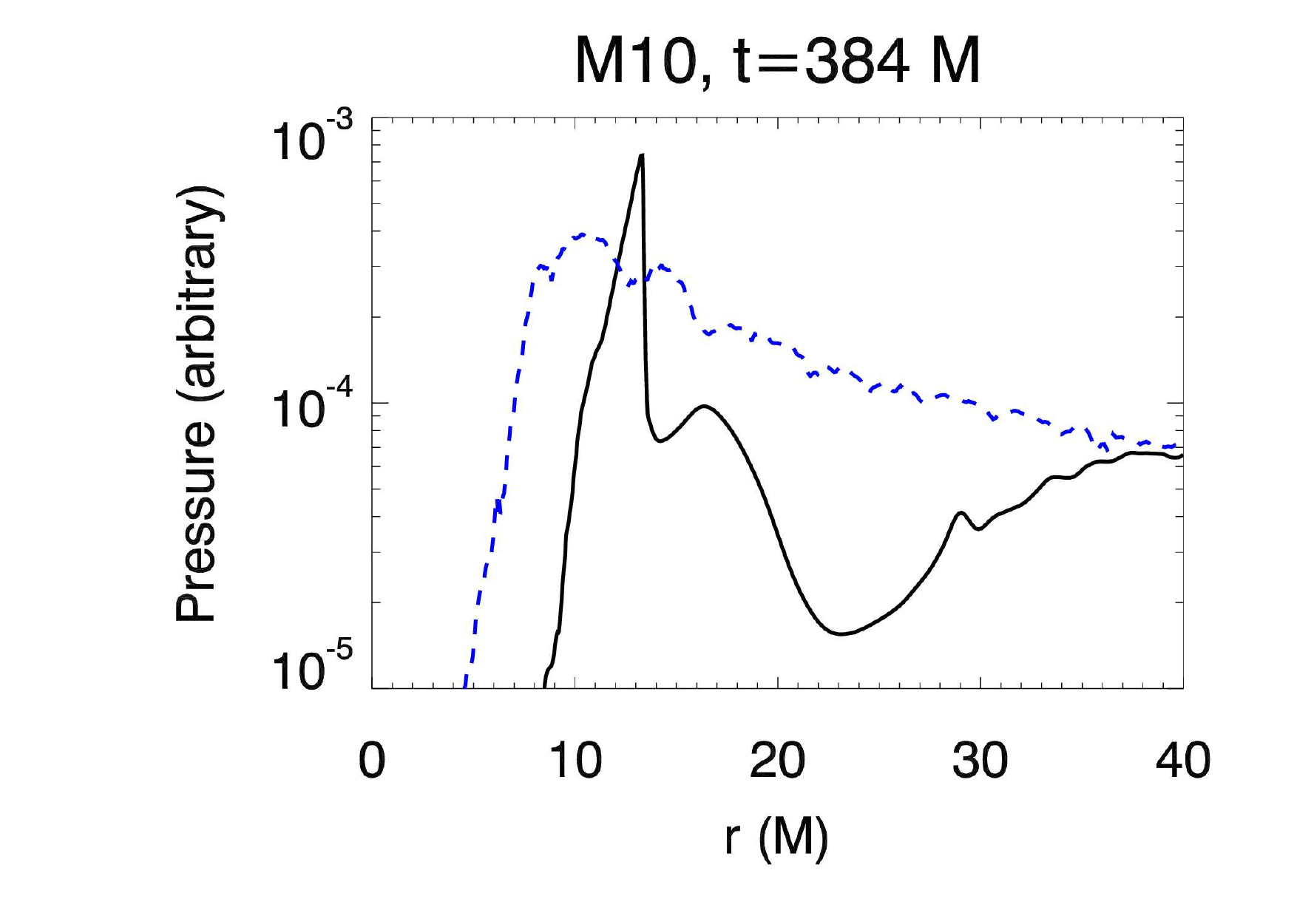}
\caption[Comparison of pressure]{Snapshots of the midplane pressure taken at $t=384~{\rm M}$ ({\it solid line}), where the central mass loss has taken place at $t=0$ ({\it dashed line}).  Each of models H5, H10, and M10 shows a peak in pressure steep enough to represent a shock driven by the oscillation of the inner disk.  In contrast, the low fraction of mass lost in model H1 produces a much weaker and broader signal.}
\label{fig:pcomp}
\end{figure*}

The baseline simulation is initialized with complete azimuthal symmetry and an aspect ratio $h/r = 0.05$, where $r=R \sin \theta$ is the cylindrical disk radius.
The initial density and pressure profiles are given by
\begin{equation}
\label{eq:dprofile}
\rho(R,\theta) = \rho_{\rm 0} \exp \left( -\frac{\cos^2\theta}{2 (h/r)^2 \sin^2\theta }\right),
\end{equation}
and
\begin{equation}
\label{eq:pprofile}
p(R,\theta)=\frac{GMR (h/r)^2 \sin^2\theta}{(R-2r_{\rm g})^2}~\rho(R,\theta), 
\end{equation}
where $\rho_{\rm 0}$ is the initial density in the disk midplane (\ie at $\theta = \pi/2$).
These forms result in an initial disk of constant midplane density and a radially decreasing temperature profile.
The initial disk is also in approximate hydrostatic equilibrium, although this is disrupted by the development of turbulence driven by the MRI. 
The initial disk velocity is entirely azimuthal with
\begin{equation}
v_{\phi}=\frac{\sqrt{GMr}}{r - 2 r_{\rm g}} \qquad v_{\rm R} = v_{\rm \theta} =0.
\end{equation}

\begin{figure*}
\includegraphics[type=pdf,ext=.pdf,read=.pdf,width=0.5178\textwidth]{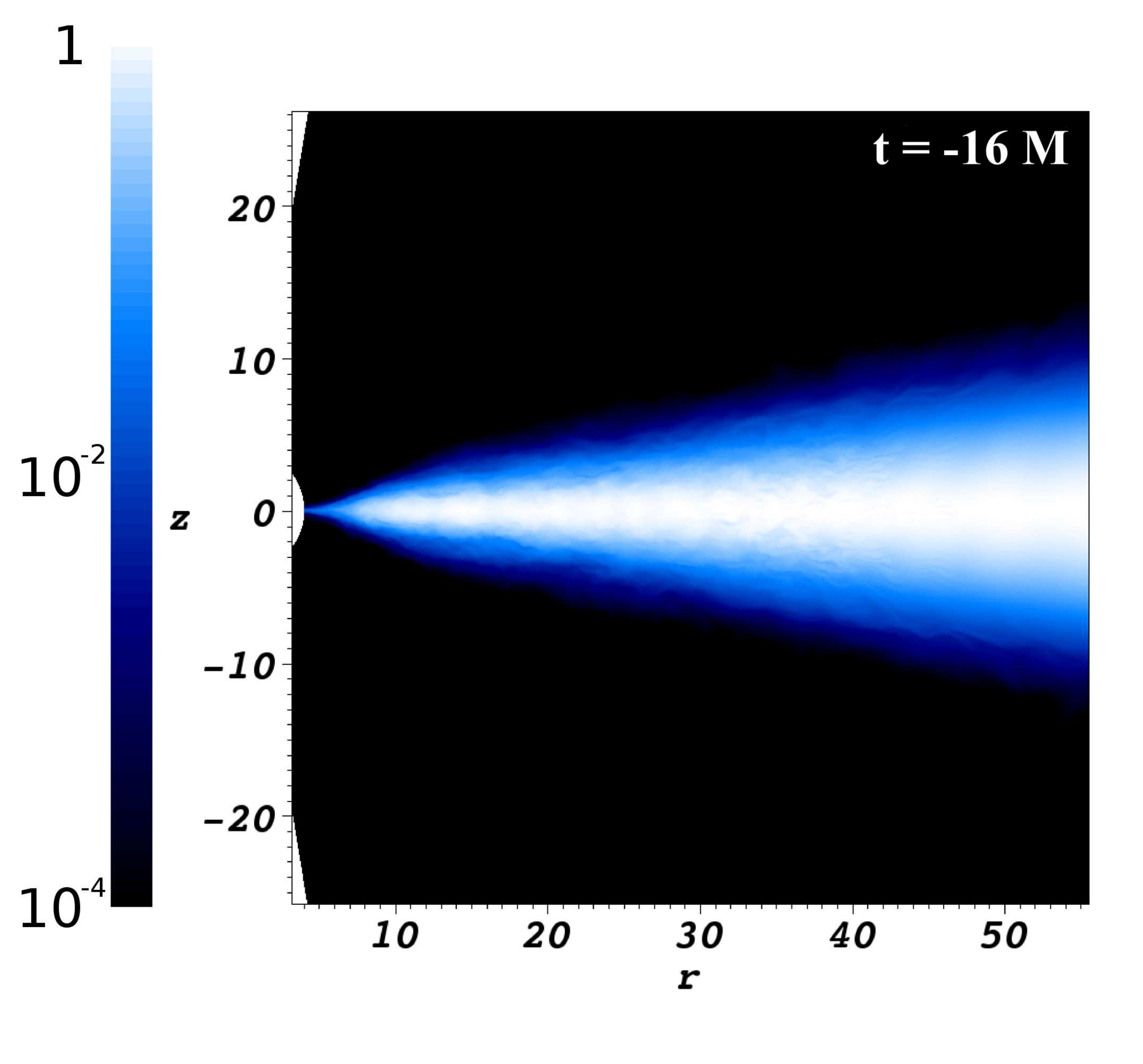}
\includegraphics[type=pdf,ext=.pdf,read=.pdf,width=0.4822\textwidth]{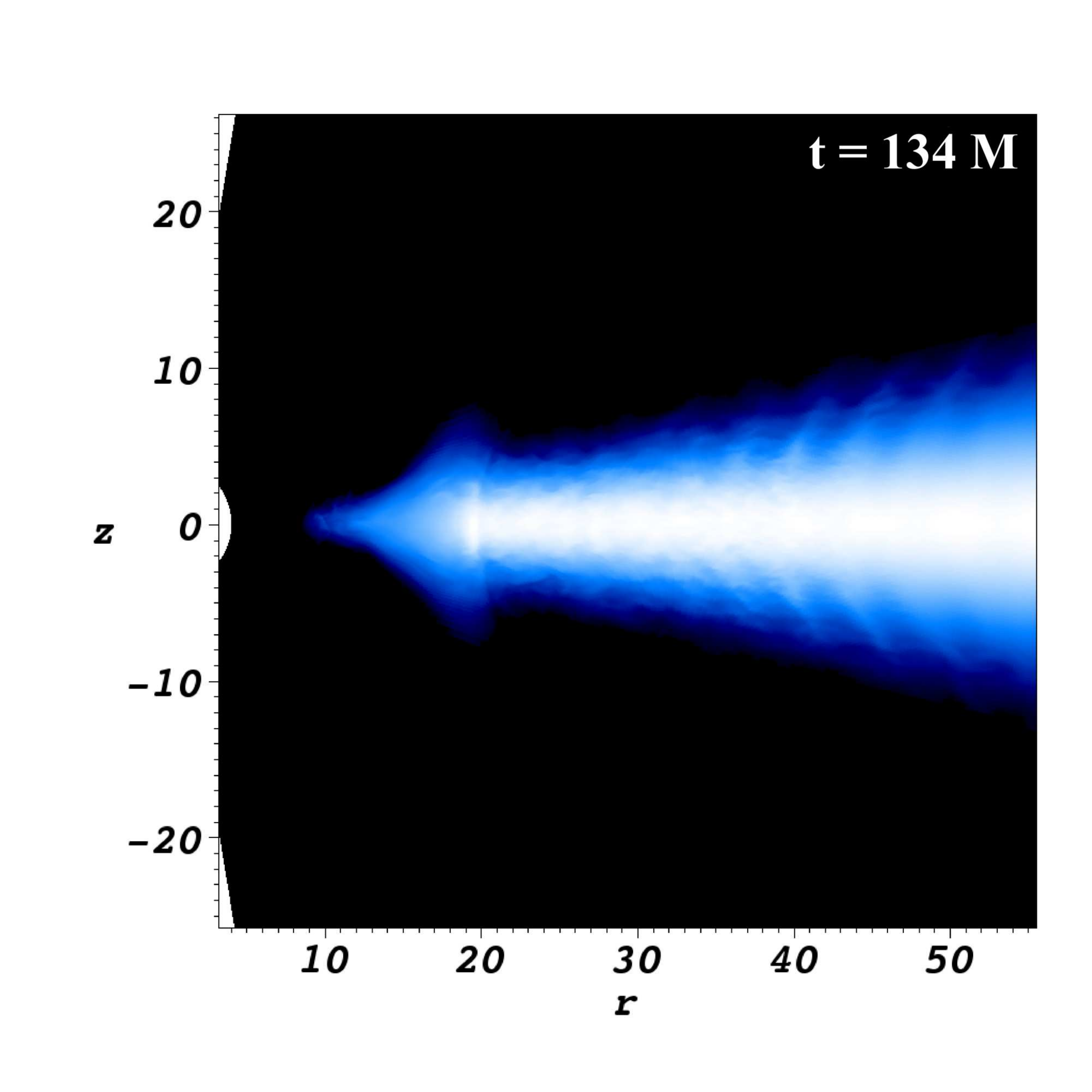}
\\
\includegraphics[type=pdf,ext=.pdf,read=.pdf,width=0.5178\textwidth]{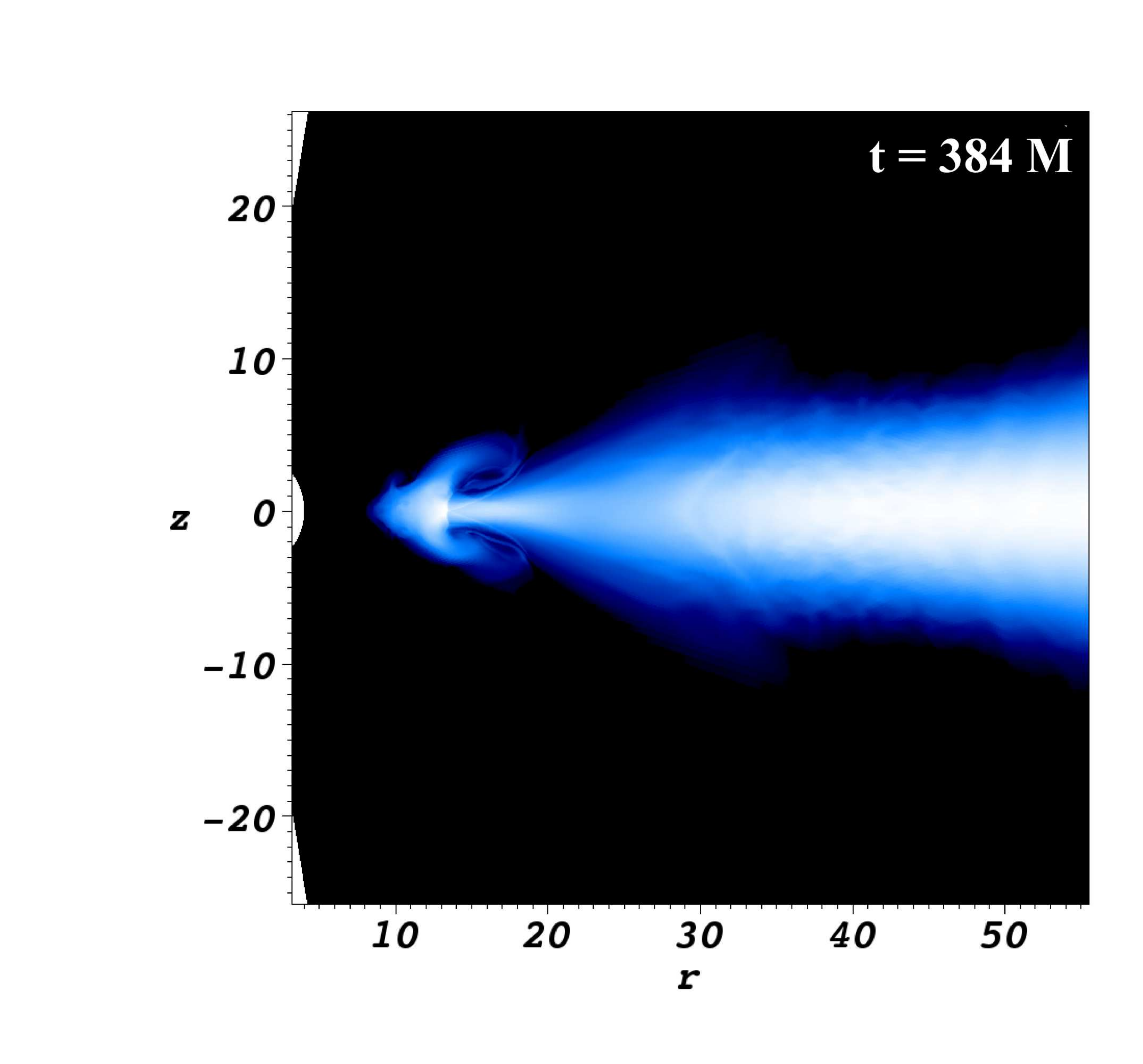}
\includegraphics[type=pdf,ext=.pdf,read=.pdf,width=0.4822\textwidth]{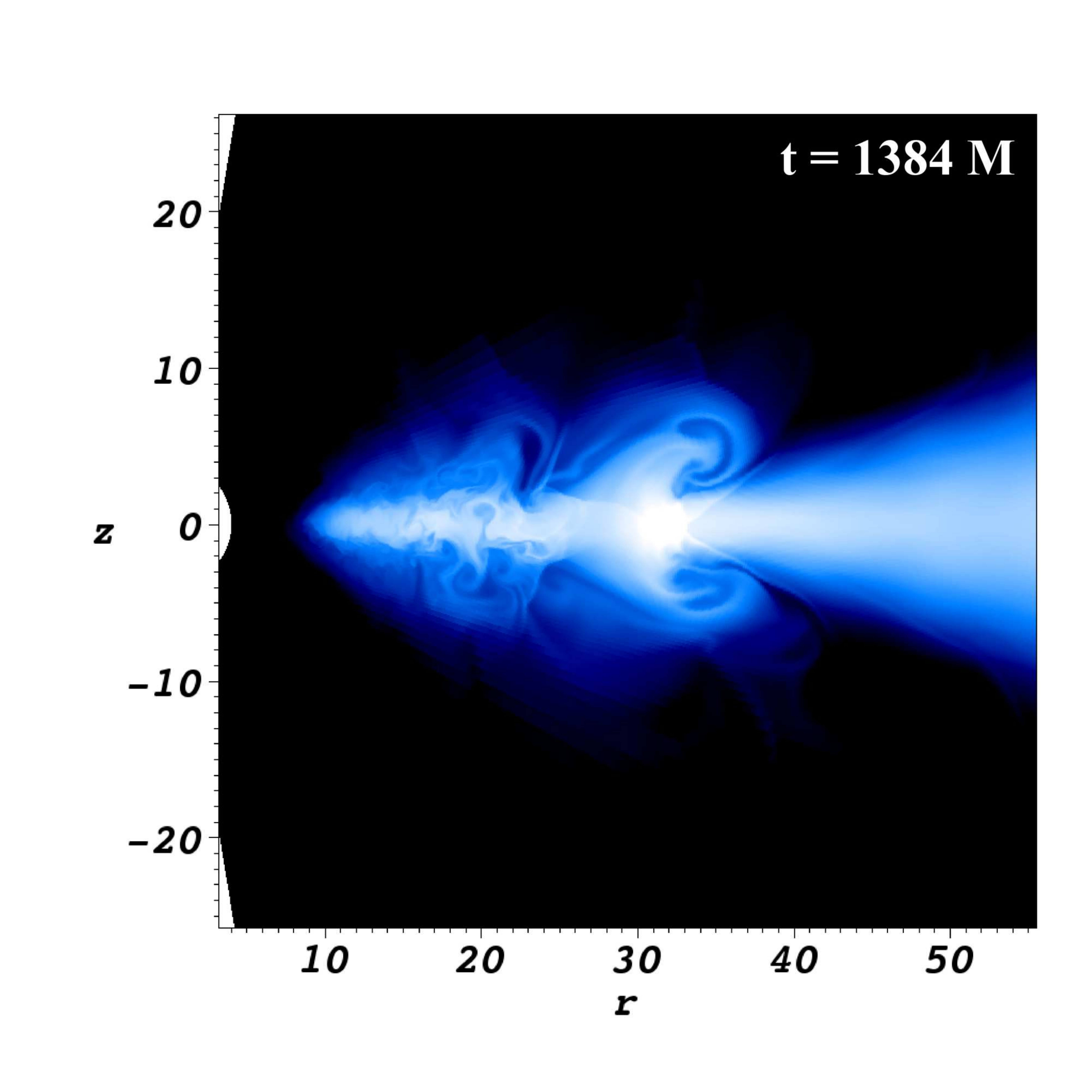}
\caption[Visualizations of H10]{Snapshots of the density in model H10 at various times.  The colorbar is logarithmic, ranging from $10^{-4}\rho_0$ ({\it dark}) to $\rho_0$ ({\it light}).  The mass-loss event takes place at $t = 0$.  The initial wave from the mass-loss event is visible at $t=134~{\rm M}$ while the stronger shock is visible at $t=384~{\rm M}$ and $t=1384~{\rm M}$.  Features such as the shocks and rarefactions propagate radially outward in the disk (from left to right in these images).}
\label{fig:vizcomp}
\end{figure*}

The initial magnetic field configuration consists of a series of field loops, each of which is contained within an $R-\theta$ plane.
This field was derived from a vector potential of the form
\begin{equation}
A_\phi=A_{\rm 0} p^{1/2} \sin \left(\frac{2 \pi r}{d_{\rm loop}(r)}\right) [d_{\rm loop}(r)]^2
\end{equation}
where $A_{\rm 0}$ is a scaling constant chosen such that the initial maximum magnetic field strength corresponds to a ratio of gas-to-magnetic pressure of $\beta \sim 3000$.
The simple function $d_{\rm loop} = 1.5 + r/20$ is included to increase the width of the field loops with increasing $r$.
The development of the MRI should be reasonably insensitive to the exact details of this field except insofar as the field has no net flux over scales larger than a few gravitational radii in the inner disk.
By the time this baseline simulation has run for $100~T_{\rm ISCO}$, the inner third of the accretion disk (\ie out to $\sim 35~{\rm M}$) has evolved to a turbulent state through the action of the MRI.
It is this state that is imported as initial conditions into our various models of mass loss from a binary merger.
For those fully 3D MHD models (denoted with an ``M''), we use exactly the output of the baseline simulation as our initial conditions.
The ``2.5D'' hydrodynamic (denoted with an ``H'') models, on the other hand, incorporate the azimuthally averaged output from the baseline simulation, neglecting the magnetic field information.
While this initial state is slightly out of equilibrium because of the omission of magnetic and/or non-axisymmetric stresses, the gross similarity between the behaviors of the MHD and hydrodynamic models suggests that this is not a significant problem. 
The only other parameter varied between models is the amount of central mass lost to gravitational waves, referred to in the model names as a percentage (\eg model ``M10'' is an MHD disk that loses 10\% of its central mass).
Each of these simulations is run for at least 20 ISCO orbital periods after the mass-loss event has taken place.
The simulations are summarized briefly in Table 1.

\subsection{Simulation Results}

We now describe the results of our simulations of accretion disks surrounding central objects that have lost mass through the prompt emission of gravitational waves.
It is worth noting that we will frequently use descriptors such as the gravitational radius and the ISCO orbital period, both of which are defined in terms of the central mass.
Rather than having to adjust our spatial and temporal reference scales after the mass-loss event, all distances and times are given in units of the {\it initial} central mass.

\subsubsection{Disk Response and the Formation of Shocks}

\begin{figure*}[t]
\includegraphics[type=pdf,ext=.pdf,read=.pdf,width=0.5\textwidth]{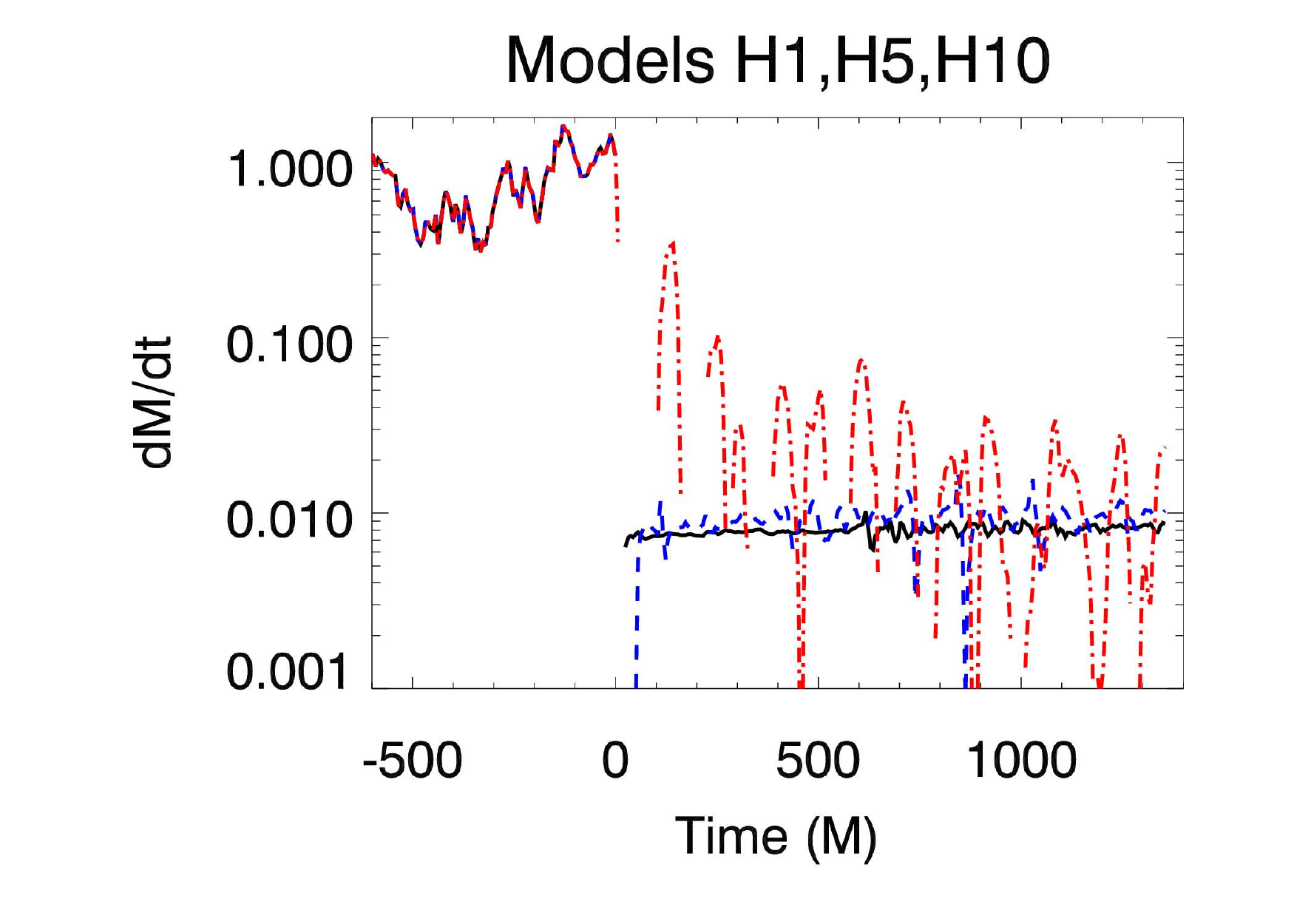}
\includegraphics[type=pdf,ext=.pdf,read=.pdf,width=0.5\textwidth]{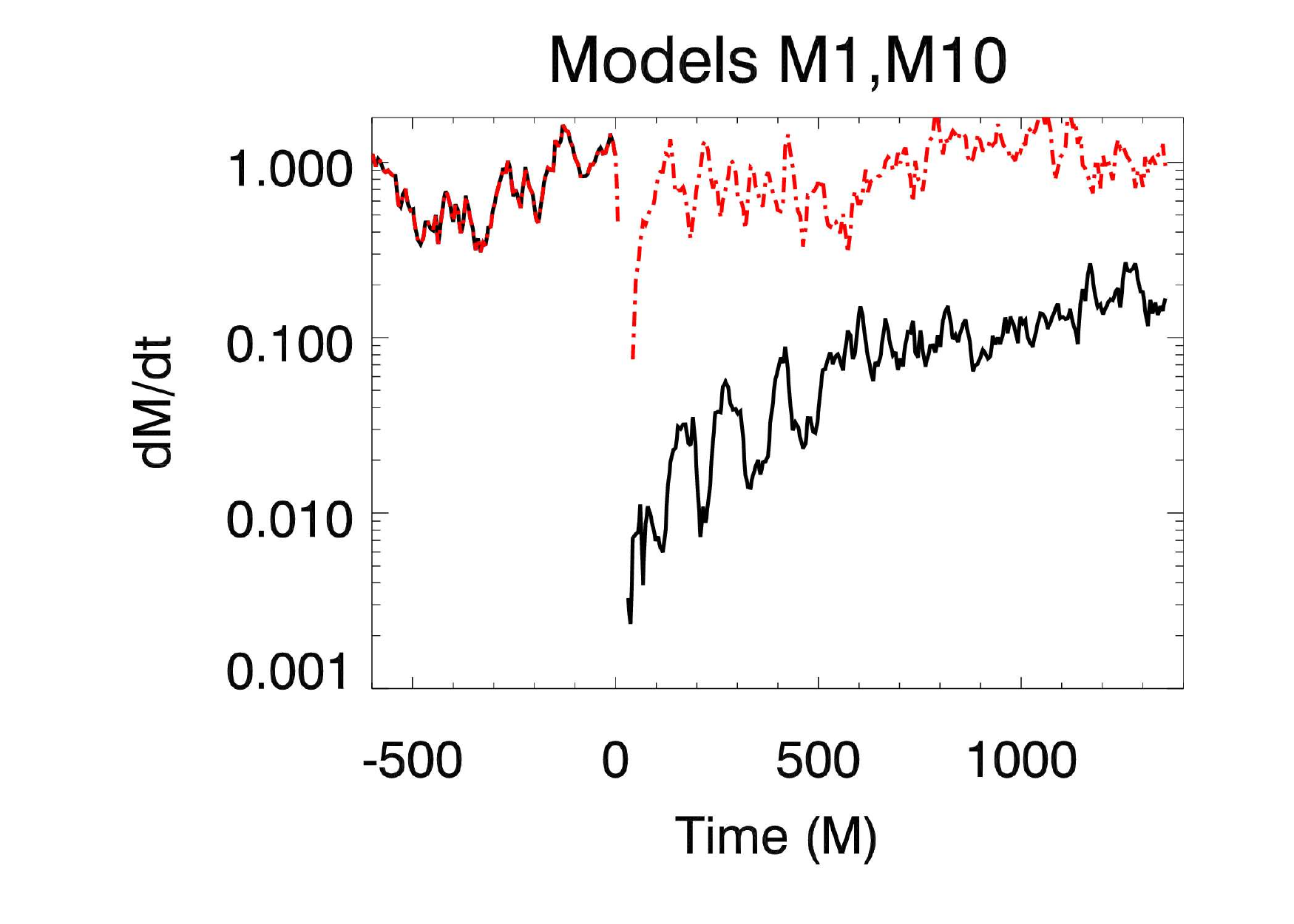}
\caption[Comparison of mdot]{Comparison of the mass accretion rate across the ISCO for hydrodynamic ({\it left}) and MHD ({\it right}) models.  Mass-loss values of $10 \%$ ({\it solid lines}), $5 \%$ ({\it dashed line}), and $1 \%$ ({\it dot-dashed lines}) are shown, with the mass-loss event taking place at $t = 0$.  In hydrodynamic disks, accretion never resumes after the mass-loss event.  In model M10, however, there is a initial drop in accretion rate, followed by a gradual resumption of accretion over the viscous timescale.  The initial accretion rate prior to the binary merger is shown at times $< 0$ for reference.}
\label{fig:mdot}
\end{figure*}

As described earlier, the basic response of any orbiting fluid element to a sudden reduction in the central mass is radial oscillation.
Since the disk material is suddenly at the pericenter of what will become its new orbit, the initial radial motion after the mass-loss event is directed outward ($v_r > 0$).
As expected, all of our simulated disks feature this generic behavior, and the amount of outward disk motion is observed to depend upon the fraction of central mass that has been lost.
Since the hydrodynamic ``H'' models and MHD ``M'' models feature very similar large-scale disk dynamics, we will focus in this section primarily on the simpler hydrodynamic models, noting differences in the MHD cases at the end of the section.

We first discuss model H10, which features the largest amount of central mass loss and therefore the most dramatic disk response.
Defining disk material by a density cutoff of $\rho_{\rm disk} \ge 0.01 \rho_{\rm 0}$, the inner edge of the H10 disk is seen to migrate out from $r \sim 4~{\rm M}$ to $r \sim 12~{\rm M}$ in just over one ISCO orbit.
The progress of the inner edge of material then stalls out at this radius while the initial perturbation from its motion continues to propagate radially outward.
The innermost edge then returns radially inward, reaching a minimum radius of approximately $r \sim 8~{\rm M}$ before beginning the next cycle of oscillation.
This first oscillation is completed in a total of roughly four ISCO orbits before the next cycle begins.
We should mention that the entire body of the disk is undergoing similar oscillations, but that the inner disk oscillation timescale is shortest and thus most easily identified.

Each of the first two inner disk oscillations in model H10 drives a signal radially outward through the body of the disk.
The detailed structures of these features change as they propagate through the disk, but they are generally identifiable at later times.
The lower-left panel of Figure \ref{fig:pcomp} shows the instantaneous midplane pressure profile for model H10 at $t= 384~{\rm M}$, approximately six ISCO orbits after the mass-loss event.
The initial disk perturbation is now seen as a small peak near $r = 28~{\rm M}$, while the second signal forms a very prominent peak at $r = 12~{\rm M}$.
The two signals are separated by a rarefaction with a pressure nearly an order of magnitude below the pre-loss pressure value (shown in Figure \ref{fig:pcomp} as a dashed line).
This rarefaction, caused by the local depletion of disk material into the regions of enhanced density and pressure, also propagates radially outward, bracketed radially by the two pressure peaks.
The steepness of the leading edge of the inner pressure feature in this figure suggests that this oscillation is driving a shock radially outward into the disk.
Similarly, Figure \ref{fig:vizcomp} shows a rendering of the disk density at four different times during the evolution of model H10.
In this figure, the initial perturbation is only easily seen at $t = 134~{\rm M}$, but the second stronger perturbation is identifiable at both $t = 384~{\rm M}$ and $t = 1384~{\rm M}$.
We also note that the disk material behind the second density/pressure enhancement can be seen expanding vertically out of the midplane, suggesting that heating is taking place.
As seen in the bottom two panels of Figure \ref{fig:vizcomp}, this heated material actually gets ahead of the midplane shock front by pushing quickly through the relatively diffuse material at higher latitudes before spiraling back into the disk.
After the first two oscillations, subsequent individual oscillation events become much more difficult to identify (see the bottom-right panel of Figure \ref{fig:vizcomp}), and the inner edge of the disk gradually settles down to a radius $r \sim 9~{\rm M}$.

\begin{figure*}[t]
\includegraphics[type=pdf,ext=.pdf,read=.pdf,width=0.5\textwidth]{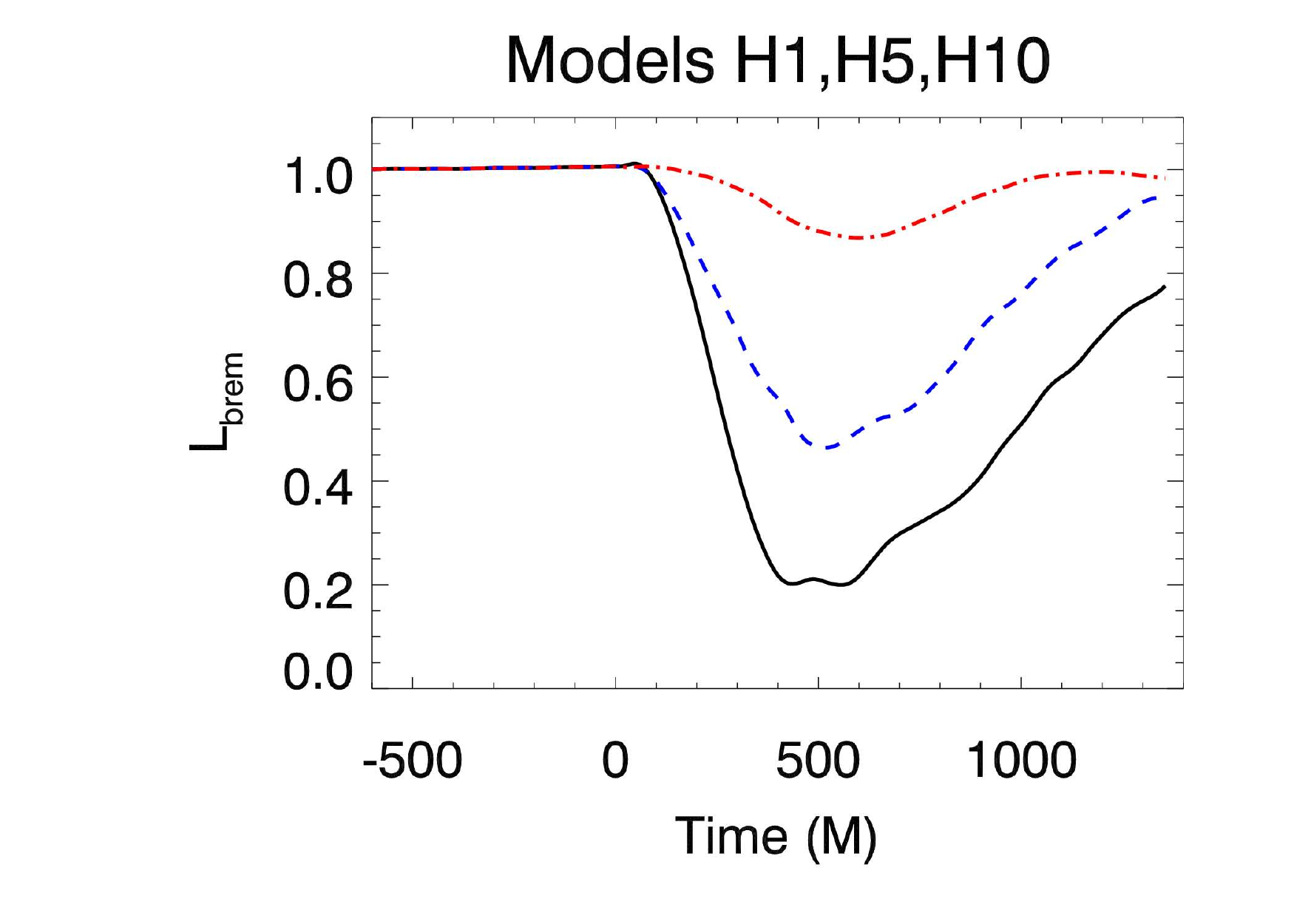}
\includegraphics[type=pdf,ext=.pdf,read=.pdf,width=0.5\textwidth]{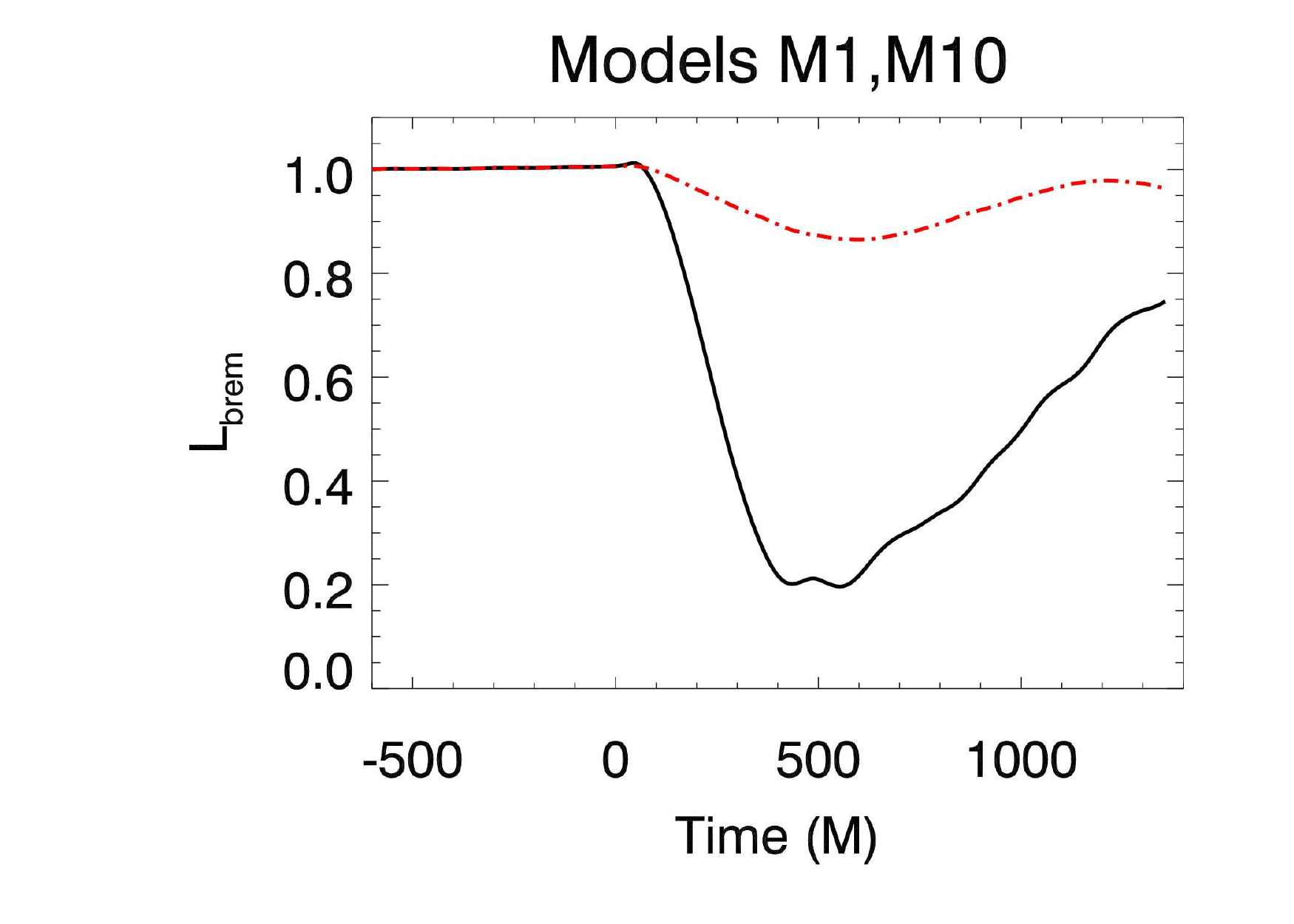}
\caption[Comparison of lbrem]{Comparison of the bremsstrahlung luminosity as measured between $R=5-35~{\rm M}$ for hydrodynamic ({\it left}) and MHD ({\it right}) models.  Mass-loss values of $10 \%$ ({\it solid lines}), $5 \%$ ({\it dashed line}), and $1 \%$ ({\it dot-dashed lines}) are shown, with the mass-loss event taking place at $t = 0$.  In both hydrodynamic and MHD disks, there is barely any increase in the luminosity.  There is, however, a drop in luminosity that depends upon the fractional mass lost, followed by a gradual recovery.  The initial luminosity prior to the binary merger is shown at times $< 0$ for reference.}
\label{fig:lbrem}
\end{figure*}

In contrast, Model H1 does not feature enough central mass loss to form shocks.
The upper-left panel of Figure \ref{fig:pcomp} shows the pressure profile of model H1, also approximately six ISCO orbits after the mass loss.
There is an identifiable feature near $r \sim 15~{\rm M}$ that corresponds to the second disk oscillation, but it is too broad to be a shock.
Moreover, it is difficult to identify the initial pressure signal, which no longer stands out above the background profile.
The inner edge of the H1 disk moves at most a few gravitational radii as a result of the mass-loss event, and this motion is simply insufficient to drive shocks into the disk.
Unsurprisingly, Model H5 features behavior that is bracketed by the other two models.
Specifically, the upper-right panel of Figure \ref{fig:pcomp} shows that H5 appears to form a shock on the second disk oscillation, but also that the initial pressure signal does not stand out.
The rarefaction between these two peaks is also visible, but reduced from that of H10.

Combined, these three simple models suggest that there is a lower limit to the amount of central mass loss capable of generating shocks in an accretion disk.
This limit is not surprising, given that the requirement for a shock to form is $v_{\rm r} > c_{\rm s}$, assuming that $v_{\rm r}$ is also an approximation of relative velocity in the system.
Rewriting $c_{\rm s} \sim (h/r) v_{\phi}$ and noting that $v_{\rm r}/v_{\phi} \approx \epsilon$ for mass-loss events, the requirement becomes $\epsilon > h/r$.
Although we have not fully explored the parameter space of mass-loss values, our simulations are consistent with this limit in that shocks only fail to form in the one model for which $\epsilon < h/r$, and we expect this simple trend to apply more generally to any thin disk aspect ratio.
As we will discuss in Section 4.1, this limits us to considering shocks only in low-luminosity systems for which $h/r < \epsilon$, which is itself much less than unity.

Finally, we point out that the 3D MHD simulations feature much of the same behavior as the hydrodynamic models.
Figure \ref{fig:pcomp} illustrates that models M10 and H10, for example, produce very similar pressure profiles at comparable times.
Specifically, model M10 clearly features a shock and an identifiable initial peak very similar to those of H10.
Likewise, the behavior of the inner disk edge is quite similar for the first few oscillations.
Model M1 is similarly akin to model H1 in that neither forms shocks.
In fact, the similarity of hydrodynamic and MHD simulations at early times is good evidence that the azimuthally averaged, non-magnetized initial conditions used in the hydrodynamic models did not induce stresses that were significant compared to those triggered by the mass loss.
There is, however, one notable difference between hydrodynamic and MHD models that reveals itself at later times.
Whereas the inner edge of model H10 settles down to $r \sim 9~{\rm M}$, for example, the inner edge of M10 gradually begins to migrate inward.
This is due to the resumption of accretion, which is accomplished in M10 through turbulence driven by the MRI.
Although model H10 features an initially turbulent disk profile, the usual means of driving turbulence are not included in the physics of the hydrodynamic models.
While this difference does not strongly affect the formation of shocks, it does affect possible observational signatures of mass loss, as we will discuss in the following section.

\subsubsection{Simulated Observations of Mass Loss}

While our simulations do not include the physics needed to assemble rigorous models of radiation production and transfer in accretion disks, we can construct simple proxies for real lightcurves.
Here, we consider the mass accretion rates and bremsstrahlung luminosities derived from our model disks to explore how these quantities reflect the central mass loss and disk dynamics described in the previous section.

First, we construct the mass accretion rate as
\begin{equation}
{\dot M} = \int_{\rm ISCO} R (-v_{\rm R})\rho dS,
\end{equation}
where the integral is taken over the surface formed by the radius of the ISCO. 
Figure \ref{fig:mdot} shows this quantity plotted for all five disk models.
As described in the previous section, the hydrodynamic models lack the physics needed to properly resume accretion after the mass-loss event.
The oscillation of the inner edge of model H1 intersects the ISCO, creating a pattern of semi-regular fluctuations, but only a negligible fraction of this inflow is accreted onto the compact object.
The MHD models, in contrast, show much more interesting and realistic accretion behaviors.
Both models M1 and M10 feature a sudden depression in ${\dot M}$ just after the mass-loss event.
This represents a decrease of over two orders of magnitude in ${\dot M}$ as accretion is temporarily disrupted.
In the case of model M1, this reduced accretion rate persists for at most two ISCO orbits ($\sim 120~{\rm M}$) before accretion across the ISCO rapidly resumes.
Model M10, however, shows a much longer-lived decrement in mass accretion rate.
After the initial drop, the accretion rate in M10 grows gradually at a rate that, if extrapolated, would match the initial accretion rate after a time $\sim 2000~{\rm M}$.
This is in fact comparable to the viscous timescale of the inner disk, as measured after the mass loss has taken place.

\begin{figure*}[t]
\includegraphics[type=pdf,ext=.pdf,read=.pdf,width=0.5\textwidth]{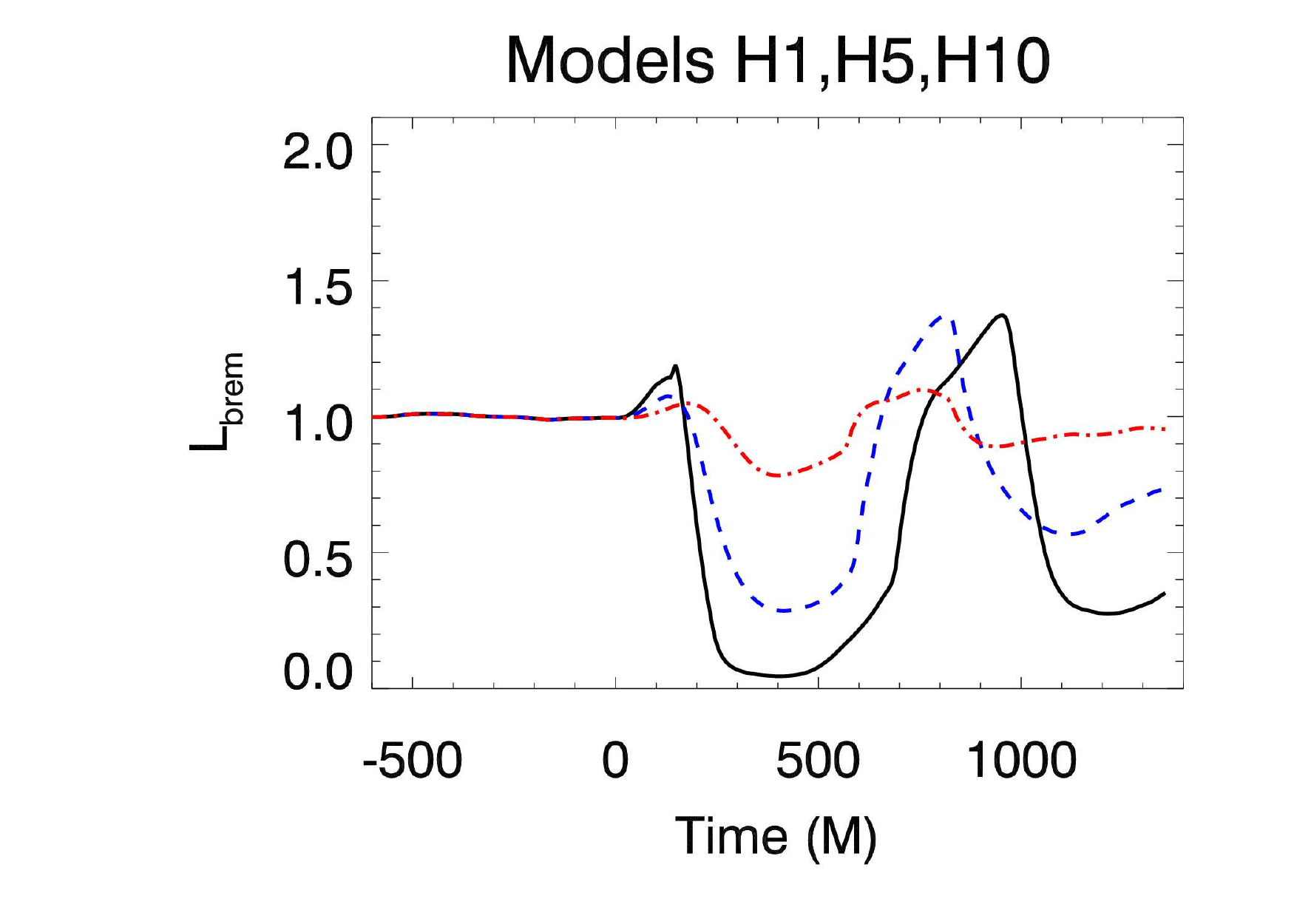}
\includegraphics[type=pdf,ext=.pdf,read=.pdf,width=0.5\textwidth]{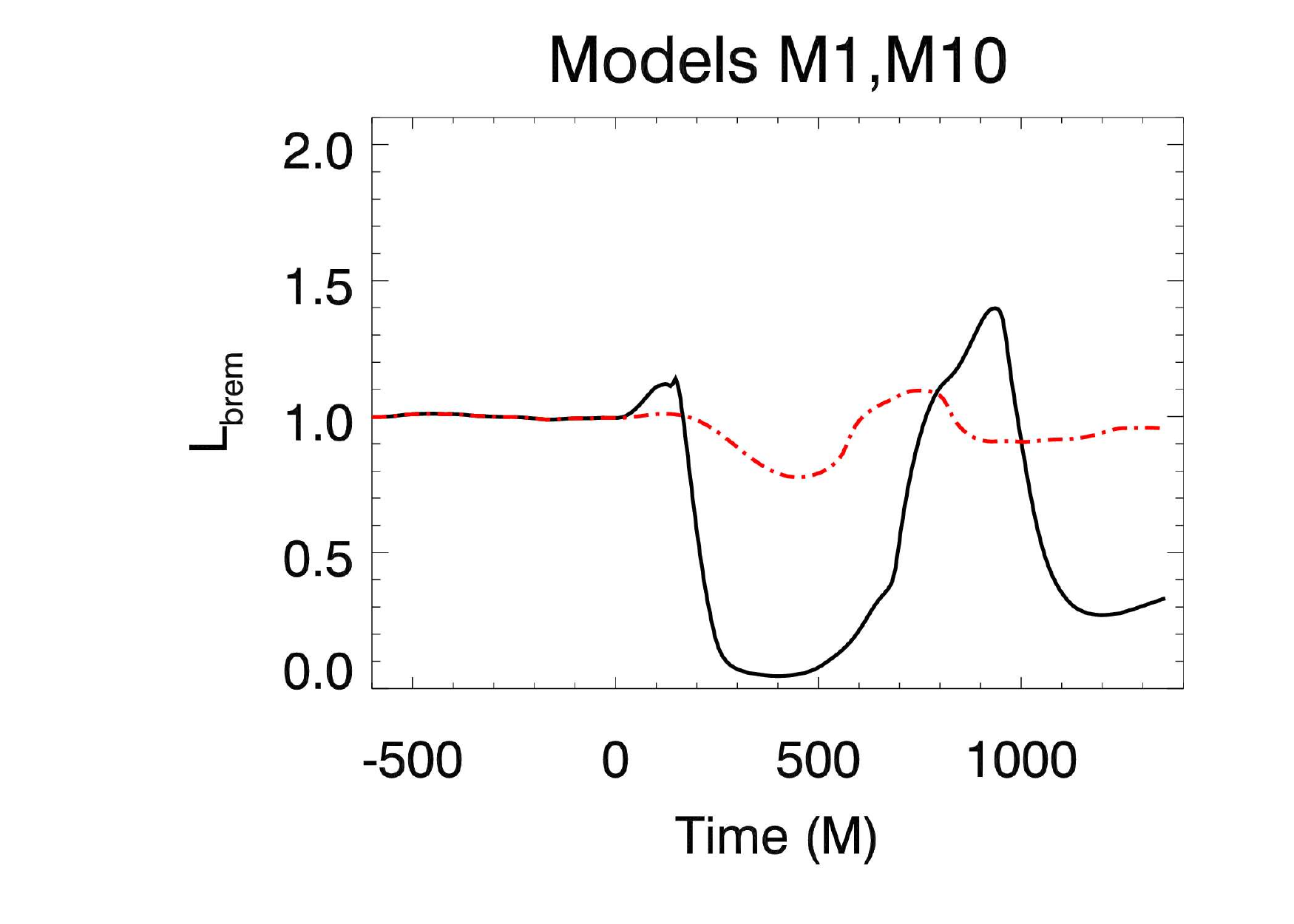}
\caption[Comparison of lbrem]{Comparison of the bremsstrahlung luminosity as measured between $R=20-25~{\rm M}$ for hydrodynamic ({\it left}) and MHD ({\it right}) models.  Mass-loss values of $10 \%$ ({\it solid lines}), $5 \%$ ({\it dashed line}), and $1 \%$ ({\it dot-dashed lines}) are shown, with the mass-loss event taking place at $t = 0$.  In both hydrodynamic and MHD disks, there is an initial increase in luminosity, followed by a drop that depends upon the fractional mass lost.  This is then followed by a sharp recovery as the shock passes through, followed by another drop.  Although there are short-lived increases in the luminosity of this single annulus, Figure \ref{fig:lbrem} shows that such features are not visible when the luminosities are summed over a larger region.  The initial luminosity prior to the binary merger is shown at times $< 0$ for reference.}
\label{fig:lbrem2}
\end{figure*}

We also examine the bremsstrahlung luminosity, which we compute from
\begin{equation}
L_{\rm brem} = \int_V \epsilon_{\rm brem} dV, 
\end{equation}
where $\epsilon_{\rm brem} = \rho^2 T^{1/2}$ is a measure of the total bremsstrahlung emissivity.
This is admittedly a somewhat naive treatment of the complex radiative processes expected in real disks, particularly those in the optically thick regime.
We argue, however, that the effect of opacity would only be to dilute or delay a radiative signature of mass loss, and that the optically thin regime constitutes an important test case.
Specifically, if our optically thin models fail to produce a significant bremsstrahlung signal, further radiative processing is irrelevant.
If, on the other hand, the bremsstrahlung emission profile proves to be a clear indicator of mass lost through a binary merger, then we must include the caveat that such emission will be subject to radiative reprocessing.
We will discuss this issue further in Section 4, where we examine in detail the observability of signals from mass loss.

Figure \ref{fig:lbrem} shows for all models the bremsstrahlung luminosity as constructed for a thick spherical shell ranging from $R=5-35~{\rm M}$.
The most interesting aspect of this figure is that none of the models feature a pronounced increase in total bremsstrahlung emission in this region.
Naively, one might have assumed that the action of shocks passing through this region would only increase the luminosity.
Instead, however, we see that the reduction in pressure and density in the rarefaction (visible in Figures \ref{fig:pcomp} and \ref{fig:vizcomp}) actually works to reduce the total bremsstrahlung luminosity, followed by a recovery as the shock passes through the region.
Simultaneously, the oscillations deplete material from the inner regions of our integration volume, also acting to lower the luminosity.
The net effect of these processes is a total bremsstrahlung luminosity that essentially never increases above the baseline level.

In order to examine only the effects of shocks and rarefactions on this disk, we show in Figure \ref{fig:lbrem2} the bremsstrahlung emission in a narrow annulus ranging from $R=20-25~{\rm M}$, well outside of the range of motion of the inner edge of the disk.
In fact, Figure \ref{fig:lbrem2} clearly shows indications of each of the pressure and density features seen in Figures \ref{fig:pcomp} and \ref{fig:vizcomp}.
Considering for the moment models M10/H10, the luminosity increases at first by about $10-20\%$ as the perturbation moves through the volume.
This is followed by a sharp $\gtorder 90\%$ net drop in luminosity as the rarefaction passes through.
The next compression causes the next luminosity increase, which can be significant for models H10, H5, and M10, all of which produce shocks.
We see, however, that shocks only increase the luminosity by at most $\sim 40\%$ above its original value.
Furthermore, this enhancement is not long-lived.
Once the next rarefaction arrives, the luminosity drops again, and it is not obvious that the net luminosity will be significantly increased over long times.
While Figure \ref{fig:lbrem2} confirms that the shocks and rarefactions are behaving as expected, we emphasize that the observed trends in luminosity are essentially local phenomena.
When the emission from several such annuli is summed, as shown in Figure \ref{fig:lbrem} for example, we see no significant increase in bremsstrahlung luminosity.
This suggests that any local increases in luminosity are constantly being offset by decreases elsewhere in the disk.

In fact, the drop in bremsstrahlung luminosity is the dominant feature in Figures \ref{fig:lbrem} and \ref{fig:lbrem2}.
As mentioned, the M10/H10 models in particular feature a $\gtorder 90\%$ reduction in luminosity, followed by a gradual recovery.
This effect is much more dramatic than the minor luminosity increase produced by the initial perturbation, as seen in Figure \ref{fig:lbrem2}.
Moreover, we see that the presence of this luminosity dip does not require shock formation.
Specifically, the H1/M1 models feature a $\sim 20\%$ reduction in luminosity and do not form shocks.
This is significant since models without shocks are no longer constrained by the $\epsilon > h/r$ requirement and may thus be related to more luminous systems.

A comparison of Figures \ref{fig:lbrem} and \ref{fig:lbrem2} also illustrates that the recovery time of the prompt bremsstrahlung luminosity drop is determined at least in part by the size of the region we choose to examine. 
The bremsstrahlung luminosity profile of M10 in Figure \ref{fig:lbrem}, for example, resembles the profile of ${\dot M}$, but the reasons for these similar profiles are quite different.
In the case of ${\dot M}$, the recovery time was determined by the physical parameters that feed into the viscous timescale.
In the case of the bremsstrahlung luminosity, Figures \ref{fig:lbrem}
and \ref{fig:lbrem2} show that the recovery time depends upon the size
of the sampling volume.  From an observational point of view, this
implies that, in the case in which the observed luminosity is
predominantly powered by accretion onto the binary, one may be able to
measure the viscous timescale of the circumbinary disk from the light
curve and thus constrain the structure of the disk.  Alternately, if
the true luminosity is dominated by emission from the disk itself
(here estimated in the form of bremsstrahlung emission), the drop in
flux may be detected as a consequence of the mass loss.  The length of
the drop in that case, however, would not reflect the viscous timescale.

\section{Observability}

The observability of signals from mass loss due to gravitational wave
emission depends on the flux and duration of the event, but also on
the angular localization possible from {\it LISA} observations.  In
this section we explore various possible signals and observing
strategies.  We focus primarily on the energetics (as opposed to detailed spectral modeling), but note that the
strongest signals are likely to be found in the X-ray band or, in some
scenarios, the radio band.  Both of these bands are dominated by
processes associated with the innermost parts of the accretion disk,
where the effects of the abrupt mass loss are the strongest.
Furthermore, they can penetrate significant columns of circumnuclear
matter that may be present.  On the other hand, ultraviolet, optical
and infrared variations are likely to be relatively weak due to
dilution by the outer accretion disk and starlight and will be much
more susceptible to absorption.  Note however that the precursor
signature in any band may be difficult to observe in Compton-thick
AGN due to obscuration.

We briefly comment upon the mechanism by which disturbances in the
accretion disk may be reflected in the observed X-ray and radio
fluxes.  X-rays are likely produced via thermal Comptonization of
softer disk photons by a corona of energetic electrons.  The energy
source for this corona is believed to be the accretion disk itself;
abrupt changes in the energetics of the disk are very likely to lead
to abrupt changes in coronal heating and X-ray emission.  Radio
emission is associated with jets formed in the inner accretion disk
or, possibly, the black hole magnetosphere.  Even in the black hole
magnetosphere scenario where the power source of the jet is ultimately
the black hole spin \citep{1977MNRAS.179..433B}, the magnetic fields
are generated by currents in the disk and hence the jet (and
associated high-frequency radio emission) will quickly respond to
changes in the inner accretion disk.  Thus, both X-ray and radio
emission are expected to drop significantly if the inner accretion
disk is disrupted or recedes and should increase as the inner accretion
disk is replenished.

As a practical matter, it is very important to recognize that AGN are intrinsically variable and have many
different spectral distributions.  For example, the narrow-line
Seyfert~1 galaxy NGC~4051 exhibits rms variation of tens of percent in
X-ray emission from $\sim 10^{-8}-10^{-3}$~Hz
\citep{2004MNRAS.348..783M}.  AGN light curves are also known to be
non-Gaussian, sometimes showing large flares \citep[\eg][]{1997ApJ...481L..15L}.  Therefore, mere detectability of the flux excess or deficit
from a given effect is not sufficient to make the source stand out
from other AGNs.  If, however, a predicted flux change happens at the
point of merger as determined by gravitational wave observations, the
probability that this is an unrelated source drops dramatically.  In
addition, such simultaneity between electromagnetic and gravitational
radiation would constrain strongly any deviation of the speed of
gravitational waves from the speed of light.

We now explore the observability of extra emission from
circularization and dissipation, and from the prompt drop in emission caused by mass
loss.

\subsection{Observation of energy from circularization}

The extra energy release due to circularization of elliptical orbits
is not likely to be observable.  The first reason is that our analytic treatment shows
that the additional luminosity due to the circularization process is
at most 10\% of the accretion luminosity.  Indeed, our MHD simulations
show that the maximum bremsstrahlung emissivity only comes back to the level achieved
prior to mass loss, and that there is not a series of peaks that could
be picked out easily from the normal MHD turbulence.
Additionally, this is independent of any radiative reprocessing that would presumably only further muddle signals of interest.

The second reason is that for shocks to develop and thus for
energy to be thermalized efficiently, our simulations show that
the fractional disk thickness needs to be $h/r<\epsilon$.  
This limits the radius of the inner edge of the disk.  To
compute this we follow \citet{2002ApJ...567L...9A}, who
note that the viscous rate at which the inner edge of a
Shakura-Sunyaev disk moves inwards is
\begin{equation}
{\dot a}_{\rm visc}=-{3\over 2}\left(h\over r\right)^2\alpha v_K
\end{equation}
where $\alpha$ is the viscosity coefficient and
$v_K=\sqrt{GM/a}$ at a semimajor axis $a$, 
whereas the rate at which the semimajor axis of a circular binary shrinks is
\begin{equation}
{\dot a}_{\rm GW}=-{64 G^3\eta M^3\over{5c^5a^3}}\; .
\end{equation}
If we define a dimensionless parameter $x$ by $a=xGM/c^2$ and set
$h/r=0.05(4\eta)$ (i.e., suppose the disk is marginally capable of
forming circularization shocks) then the viscous and binary shrinkage
speeds become
\begin{equation}
\begin{array}{rl}
{\dot a}_{\rm visc}&=-0.024\eta^2(\alpha/0.4)x^{-1/2}c\\
{\dot a}_{\rm GW}&=-12.8\eta x^{-3}c\; .\\
\end{array}
\label{eq:radial}
\end{equation}
Setting the two equal gives
\begin{equation}
x_{\rm min}\approx 20(4\eta)^{-0.4}(\alpha/0.4)^{-0.4}\; .
\end{equation}
In these equations we scale by a viscosity parameter $\alpha=0.4$,
which is the maximum of the range inferred from observations
summarized by \citet{2007MNRAS.376.1740K}.  Therefore, in units of
gravitational radii, the closest the disk could get to the merging
binary is about $20~{\rm M}$, and this occurs for equal-mass black
holes.  In reality, the inner edge will be farther away, because for
comparable-mass black holes the disk is truncated at about double the
binary semimajor axis.  This implies a minimum of about $40~{\rm M}$, or a
bit further because at that larger radius, ${\dot a}_{\rm visc}$ is
reduced from the value we calculated above.  Offsetting this somewhat is
that the disk will drift in while the binary spirals in, so we will
take $r_{\rm inner,min}=30~{\rm M}$ as an estimate of how close the inner
edge of the disk can be at merger and still produce shocks after mass
loss.  This implies a radiation efficiency of
$E=1-|u_t|=1-(x-2)/\sqrt{x(x-3)}=0.016c^2$ for a Schwarzschild
spacetime, which should be a good approximation at this distance.

To estimate the maximum mass accretion rate we use the disk
solution of \citet{1973A&A....24..337S}.  In the inner
radiation-dominated portion of their solution (valid for all
the radii of interest here), they find a disk half-thickness of 
\begin{equation}
h=3.2\times 10^6~{\rm cm}~(M/1~M_\odot)({\dot M}/{\dot M}_E)[1-(r/6M)^{-1/2}]
\end{equation}
where ${\dot M}_E=L_E/(0.06c^2)$ and 
$L_E=1.3\times 10^{38}(M/M_\odot)$~erg~s$^{-1}$ is the Eddington luminosity.
In this region the half-thickness depends only weakly on radius.
At $r=30M$, therefore, $h/r\approx 0.4({\dot M}/{\dot M}_E)$.
For 
$\epsilon\approx 0.05$ the requirement $h/r<\epsilon$
means ${\dot M}/{\dot M}_E<0.125$.  The maximum disk luminosity is
therefore $0.016c^2{\dot M}=0.125\times(0.016/0.06)L_E=0.03L_E$.

To be specific, consider an equal-mass binary of total mass $2\times
10^6~M_\odot$ accreting at this rate, at a redshift of $z=1$ and thus
a luminosity distance of $d_L = 6.6$~Gpc (assuming a flat universe with
matter parameter $\Omega_m=0.27$ and Hubble parameter $H_0 = .71$).  The maximum observed flux is then
$F_{\rm max}=1.5\times 10^{-15}$~erg~cm$^{-2}$~s$^{-1}$. According to
our earlier estimate of Newtonian shocks, the flux variation due to
mass loss would be a factor of $\sim 10$ lower than this.  The flux
variation would therefore be just a few$ \times 10^{-16}$~erg~cm$^{-2}$~s$^{-1}$, so even if {\it all} of the radiation emerges in
X-rays it is only comparable to the dimmer sources in the 1Msec
Chandra Deep Field \citep{2002ApJS..139..369G}.  The duration of this
effect would only be roughly the orbital time at 30~{\rm M}, or $2\pi\times
30^{3/2}\approx 1000~{\rm M}$, which for $M=2\times 10^6~{\rm M_\odot}$ is only
about $10^4$ seconds.  

The most capable X-ray observatory currently planned that may operate
contemporaneously with LISA is the {\it International X-ray
Observatory} (IXO).  However, even for IXO (and temporarily putting
aside issues of angular localization and field-of-view; see \S4.3),
the flux limit of a $10^4~{\rm s}$ integration is $\sim
10^{-15}$~erg~cm$^{-2}$~{\rm s}$^{-1}$.  So even if this one-orbit variation
were distinguishable from the natural fluctuations caused by MHD
turbulence, it is too faint to be practically detectable.  

Although we have focused on the case in which $h/r \sim \epsilon$, we should mention that smaller $h/r$ ratios will generally allow for the development of more powerful shocks.
Unfortunately, such disks also have correspondingly larger inner edge radii and lower characteristic luminosities, so the prospects for detection are not necessarily improved.
Finally, there is the strong possibility that accretion onto the binary system will not be perfectly steady.
As discussed by, for example, \citet{1999MNRAS.307...79I}, \citet{2002ApJ...567L...9A}, \citet{2005ApJ...622L..93M}, \citet{2008ApJ...672...83M}, and \citet{2009arXiv0904.1383H}, torques induced by the binary are likely to modify the structure of the inner disk.
While the dominant effect is expected to be a clearing out of disk material within approximately twice the binary orbital radius, as we have assumed above, it is also likely that such torques will modify the density structure of the inner disk and, as discussed in \citet{2002ApJ...567L...9A} and \citet{2008ApJ...672...83M}, drive spiral waves into the disk body.
Although we have not explicitly modeled such features, it seems unlikely that they would achieve a signal of the energy release from circularization stronger than that which occurs in a steady disk. 
This particular electromagnetic signature of binary merger therefore remains a poor candidate for detectability.

\subsection{Observation of prompt drop in flux}

In contrast, our simulations show that the drop in mass accretion rate
caused by the outward movement of the inner edge of the disk can last
for the viscous time at that radius. Because this effect does not
require formation of shocks, the $h/r < \epsilon$ requirement can be
relaxed.  Therefore, the abrupt reduction in mass accretion could be
observable in the form of a prompt drop in flux even in systems close to
the Eddington luminosity. In these systems the viscous time is short
enough that the inner edge of the disk can keep up with the binary all
the way to the point of dynamical instability (note that if we set
$h/r\sim 1$ and $\alpha=0.1$ and solve as in
equation~(\ref{eq:radial}), we find that the disk stays with the
binary until $r\approx 6~{\rm M}$). This increases the maximum flux for
systems by a factor of 30, or up to $F_{\rm max}\approx 5\times
10^{-14}$~erg~cm$^{-2}$~s$^{-1}$, compared to the flux estimate for a
geometrically thin disk in \S~4.1.  The recession of the disk edge
following the black hole merger (and subsequent mass loss) will lead
to order unity decrease in the total flux, i.e., a transient that is
$\sim 200$ times stronger than the luminosity effect expected from
circularization.  At such high accretion rates, the viscous time is
$t_{\rm visc}\sim t_{\rm orb}/\alpha$, implying that the flux deficit
will last for $600-2000~{\rm M}$ depending on the location of the inner edge.
For a total binary mass of $\sim 2\times 10^6~{\rm M_\odot}$ and a redshift
of $z=1$ this translates to $\sim 1-4\times 10^4$~seconds.

Of course, since we are now considering the cessation and then
resumption of accretion, we need to include the bolometric correction
factor linking this total flux with that in the X-ray band.  For these
small mass systems, we will assume that $10\%$ of the total flux
emerges in the X-ray band.  Thus, the signal we seek is the following:
an AGN with an X-ray flux of $F_X\sim 5\times
10^{-15}$~erg~cm$^{-2}$~s$^{-1}$ which abruptly turns off at the
moment of merger and then re-emerges on timescales of $\sim 1-4\times
10^4$~seconds.    

Localization of the source on the sky prior to the coalescence of a
supermassive black hole binary can be achieved if there is a clear
electromagnetic signature associated with the host galaxy. Examples of
precursor signatures previously discussed in the literature include
the disturbed morphology of the host galaxy and starburst, as well as
the precursor accretion episodes, and are typically expected to occur
in case of binaries heavier than $\sim 10^7 {\rm M_\odot}$
\citep{2002ApJ...567L...9A,2006MNRAS.372..869D}.   However, less-massive 
binaries (with masses $<5 \times 10^6 {\rm M_\odot}$) are not expected to
produce strong precursor signatures, for several reasons.  First, in
the time required for binary black holes to coalesce (which could be
up to a few Gyr), any gas trapped within the binary orbit would have
been consumed, reducing the probability that the mass accretion rate
is close to Eddington.  Second, in this time it is also expected that
the starburst produced by the galactic merger would have run its
course, and in addition it is unclear whether there are any remaining
observable signs of a disturbed galactic morphology from the merger.
Low-mass binaries are therefore expected instead to be {\it followed}
by an electromagnetic afterglow rising on the viscous time scale of
the circumbinary disk
\citep{2005ApJ...622L..93M,2006MNRAS.372..869D}.

We do note that even low-mass binary systems may be identifiable
prior to coalescence and mass loss due to the {\it increasing radiative
efficiency} as the inner radius of the disk decreases while the binary
separation decays. For example, consider an equal-mass binary at a
separation of $10M$, where $M$ is the total mass of the binary.
Numerical simulations and simple estimates show that the inspiral time
from this point is approximately $1000M$ (e.g.,
\citealt{2007PhRvD..75l4024B}). At lowest order the
inspiral time scales as the fourth power of the separation
\citep{1964PhRv..136.1224P}, hence a binary with a separation of $20M$
would spiral in within $1.6\times 10^4$~M, which for a $M=2\times
10^6~{\rm M_\odot}$ binary at redshift $z=1$ would appear to us as about
four days.  If the inner edge remains at twice the binary semimajor
axis for the duration of the inspiral, this implies that the radiative
efficiency and hence the luminosity may roughly double in that time.
This is potentially observable, although we again caution that AGN are
known to have flux variations of several tens of percent over hours to
days (e.g., \citealt{2004MNRAS.348..783M}).  We are also assuming that
the luminosity from accretion onto the individual black holes is small
compared to the luminosity from the circumbinary disk.  This is
expected in the context of our model, where the density of the gas in
the innermost region of the disk is very low in comparison to the disk
mid-plane and the disk is assumed to be axisymmetric.

In a realistic scenario the circumbinary disk will not be perfectly
axisymmetric, and the black holes will not have equal mass.  It is
therefore possible that the rate
at which the gas flows from the inner rim of the disk into the low
density region around the binary may be significant. Some of
this gas will be accreted by the black holes
\citep{2008ApJ...682.1134H}, while the remainder develops eccentric
orbits, and may collide with the inner rim of the circumbinary disk or
leave the binary system in form of high velocity outflows
\citep{2002ApJ...567L...9A,2008ApJ...672...83M}.  Accretion on one or
both members of the binary and gas stream collisions with the disk may
give rise to {\it quasi-periodic outbursts} on the orbital time
scale. In the case of the binary example considered above this translates
into a variability time scale of about $\sim 3$~hr in the frame of the
observer. The variability time scale is expected to decrease monotonically
as the orbit of the binary decays. In order for
quasi-periodic outbursts to stand out above the circumbinary disk
emission, their luminosity would have to be at least at the level of
$F_{\rm max}$. Moreover, if a correlation can be demonstrated 
between pre-coalescence quasi-periodic outbursts and gravitational wave
emission, this would be a telling sign that the electromagnetic
signature is related to the binary orbit \citep{2008ApJ...684..870K}.

We emphasize that the observation of a drop in mass accretion rate, unlike the situation described in Section 4.1, assumes the maintenance of accretion during the inspiral period.
If the inner edge of the accretion disk is stalled at large radii ($r \sim 40-100 M$), as is expected for geometrically thin disks \citep[\eg][]{2002ApJ...567L...9A,2005ApJ...622L..93M}, then the associated dropoff in accretion would not directly reflect the merger event.
As we have noted, however, systems that accrete near the Eddington rate should naturally feature disks with aspect ratios of order unity.
Such disks should follow the binary system to small radii and, in the absence of appreciable torques, undergo accretion on to the central object(s).
If the binary torques on the disk are sufficient to substantially retard accretion, however, the associated accretion behavior would again fail to reflect the merger itself.
It is only for the case in which accretion is modified but not completely prevented, such as the quasi-periodic outburst case described above, that a torqued disk could feature accretion behaviors directly indicative of a binary merger event.

On the other hand, a dropoff in emission from the {\it body} of the disk, such as that discussed in Section 3.3.2, requires no constraining assumptions about accretion.
In that section we used bremsstrahlung as a simple, generic proxy for disk emission and found that the luminosity of the inner $r \sim 30 M$ of the disk was reduced by a factor of $2~(5)$ for $\epsilon = 0.05~(0.10)$.
In this scenario, an important caveat is that the inner portions of real disks are probably optically thick and would therefore reprocess this radiation.
To address this issue, we estimate the approximate radiative diffusion time $t_{\rm diff} \sim h\tau/c$, where $h$ is again the disk half-thickness and $\tau$ is the optical depth, in this case due to Thomson scattering.
Again applying the model of \citet{1973A&A....24..337S} for the radiation-dominated inner portion of an accretion disk,
\begin{equation}
\tau = \frac{1.8~(r/6M)^{3/2}}{\alpha({\dot M}/{\dot M_E}) [1-(r/6M)^{-1/2}]} .
\end{equation}
Combining this with the expression for disk height given in (19), we find that 
\begin{equation}
t_{\rm diff} \sim 4.4 \times 10^{4}{\rm s}~\left(\frac{0.1}{\alpha}\right)\left(\frac{M}{2 \times 10^6 ~M_\odot}\right)\left(\frac{r}{30M}\right)^{3/2}.
\end{equation}
For our fiducial values of $\alpha = 0.1$, $M = 2 \times 10^6 M_\odot$, and $r=30 M$, $t_{\rm diff} \sim 12~{\rm hours}$, which, at a redshift of $z=1$, would appear to us to take a day.
This is only a few times the local orbital period and rapid enough that scattering should not significantly inhibit the prompt emission of radiation from the disk body.

We note that the example flux estimates in this section were
calculated for a binary of mass $2\times10^6~{\rm M_\odot}$.  More
massive binaries, however, would coincide with more luminous EM
sources and may thus be easier to localize on the sky.  In addition,
the flux deficit in the light curves of these objects would last for a 
proportionally longer time.
If such systems are too much more massive than our fiducial value, however, they will begin to fall outside of the LISA detection band.

Although the uncertainties in the sub-parsec structure of a circumbinary
region are reflected in the uncertainties of the associated signatures, it
is plausible that the EM signatures are bracketed by the two simple
scenarios described above.
We therefore expect that the prompt drop in flux due to the mass loss
is detectable only if the proper binary system can be identified. To
determine the likelihood of this, we now consider the probable error
boxes for such systems and requirements for the electromagnetic sky
surveys of the future.

\subsection{Angular localization of coalescences}

A number of researchers have estimated the angular uncertainty of
black hole coalescences from LISA data \citep[\eg][]{1998PhRvD..57.7089C,2000AIPC..523..403S,2002PhRvD..65f2001M,2002MNRAS.331..805H,2004PhRvD..70d2001V,2005PhRvD..71h4025B,2005ApJ...629...15H,2006PhRvD..74l2001L,2008ApJ...677.1184L}.
The size of the final
error region, after inspiral, merger, and ringdown, will depend on the
direction, redshift, black hole masses, and other factors, but also on
whether the spin axes of the black holes are aligned with each other
and with the orbital axis.  If they are not, then
\citet{2006PhRvD..74l2001L} show that precession in the final day of
coalescence between typical supermassive black holes can reduce the
major and minor axes of the error ellipses by factors of a few each.
With this reduction the median major axis at redshift $z=1$ for
primary masses in the $\sim 10^5-10^7~{\rm M_\odot}$ range is $\sim
15^\prime-45^\prime$ and the median minor axis is $\sim
5^\prime-20^\prime$.  However, models of supermassive black holes in
gas-rich environments (i.e., the ones most favorable for high
accretion rates and detectability) suggest that when the holes are
still hundreds of parsecs from each other and thus their dynamics are
dominated by the gas instead of each other, gas torques will align the
spins with each other and with the orbital axis
\citep{2007ApJ...661L.147B}.   If the axes are
not aligned then the final error box near the point of merger is
typically a few hundredths of a square degree at redshift $z=1$.  If
the axes are aligned, then the lack of precession degrades this to a
few tenths of a square degree
\citep{2006PhRvD..74l2001L}.  

Whether or not the axes are aligned, a day before merger the error
region is likely to be in the square degree range.  Given the
realistic response time of an X-ray observatory such as IXO, this is
the error box relevant to X-ray searches for electromagnetic
counterparts.

How easily could such a source be observed?  Continuously monitoring a
whole LISA error ellipse on the sky in X-rays does not appear to be a
possibility since no X-ray telescope is currently planned that has a
field-of-view of several square degrees and a flux sensitivity of
$10^{-15}$~erg~cm$^{-2}$~s$^{-1}$ over $\sim 10^4$ seconds of exposure
time. A more successful strategy would be to observe a precursor
electromagnetic event, identify a source, and monitor it throughout
the coalescence.  The {\it International X-ray Observatory} has a
planned field of view of $18^\prime$ (diameter) and a flux sensitivity
of $10^{-15}$~erg~cm$^{-2}$~s$^{-1}$ for exposures of $10^4$ seconds.  Thus,
approximately 15 frames would be required to cover a square degree,
which at $10^4$ seconds for each pointing amounts to approximately two days. At
the planned sensitivity of the IXO the X-ray outbursts could be
detected for binaries with orbital periods of a few hours and longer,
in the frame of the observer. At least two subsequent snapshots of the
same sub-field of view are necessary in order to detect such an
outburst. The monitoring of the precursor increase in luminosity of
the disk requires a similar strategy (multiple exposures of the same
field of view), except the baseline between the observations of the
same sub-field can be longer than in case of the outbursts and at most
a few days apart.  Note, however, that in cases for which the LISA
error ellipse is smaller than the IXO field of view, the X-ray
observatory may immediately focus on that part of the sky without
needing to sweep over a larger area.
If a candidate with an EM precursor were found, continuous monitoring by IXO would be able to detect the flux drop-out
and subsequent recovery predicted by our models.

Concurrently, the LISA error box could be monitored in the radio band.
The Square Kilometer Array is projected to have a one square degree
field of view and a $1\mu$Jy sensitivity for a one hour exposure.  At
10~GHz, $1\mu$Jy is $10^{-19}$~erg~cm$^{-2}$~s$^{-1}$, so if more than
a fraction $\sim 10^{-5}$ of the total emission is in radio waves this
will be detected easily, as will the change in emission during the
preceding inspiral.

If it is possible to identify a counterpart to within a few
arcseconds, then further followup observations could be pursued in
other wavebands with yet higher angular resolution. We note that if
few-arcsecond positions are possible then since magnification
variations from weak gravitational lensing are dominated by scales
above tens of arcseconds \citep{2003ApJ...585L..11D} it might in
principle be possible to construct detailed shear maps of the region
and get a moderately refined estimate of the actual magnification,
although this would require a high surface density of background
galaxies, in excess of $10^6$~deg$^{-2}$.  If this procedure yields
better than the usual $\sim 10$\% precision limited by lensing, as discussed by \citet{2007ApJ...658...52J}, these
sources would serve as very precise cosmological probes.

\section{Conclusions}

We have presented a combined analytic and numerical analysis of the
effects on a circumbinary accretion disk of central mass loss caused
by prompt gravitational wave emission from a binary black hole inspiral
and merger.  Our significant results are:

1) The primary effect of abrupt mass loss caused by the merger of
supermassive black holes will be a rapid drop in the mass accretion
rate across the inner edge of the disk.  If the mass accretion rate
is close to Eddington and the inner edge can extend without significant disruption close to
the innermost stable circular orbit in the last stages of binary
inspiral, then this drop could result in large changes in the X-ray
spectrum due to, e.g., dramatically reduced Comptonization from
hot optically thin gas inside the ISCO.  The flux changes from
this effect would be detectable with current instruments
if the source were located precisely enough, but a full survey of
the likely error ellipses would require an instrument with a field
of view of at least several tenths of a square degree and sensitivity
in a $\sim 10^4$ second exposure to fluxes below 
$\sim 10^{-15}$~erg~cm$^{-2}$~s$^{-1}$.  

2) The expected enhancement of disk emission following a binary merger due to the circularization of elliptical orbits is not likely to be observed.  
While shocks are identifiable in our simulations, they contribute little to the bremsstrahlung emissivity of the disk.
Moreover, such shocks are separated by rarefactions that act to reduce the brightness of the disk.

3) In addition to serving as probes of cosmology and general relativity, observations of EM signatures from merging binary black holes have the potential to tell us much about the inner structure of accretion disks.  
If the observed emission from such systems is tied to the accretion rate in particular, the duration of the observed flux deficit can be used to estimate the viscous timescale of the disk.

\acknowledgments
SMO, CSR, and MCM acknowledge the support of NSF Grant AST 06-07428.
MCM also acknowledges NASA ATFP grant NNX08AH29G.
TB thanks the UMCP-Astronomy Center for Theory and Computation Prize Fellowship program for support.
JDS acknowledges the support of the Chandra Postdoctoral Fellowship Program.
We also thank the NCSA at the University of Illinois in Urbana-Champaign for developing ZEUS-MP.
Some of the simulations described were run on the ``Deepthought'' High Performance Computing Cluster maintained by the Office of Information Technology at UMCP.
We also acknowledge the helpful comments made by the anonymous referee.

\bibliography{refs}

\end{document}